\documentclass[prb,twocolumn,aps]{revtex4-1}
\usepackage{graphicx}     
\usepackage{amsmath,amssymb,dsfont}

\usepackage{xcolor}
\definecolor{grassgreen}{rgb}{.4,0.6,.10}
\definecolor{darkblue}{rgb}{0,0,.65}
\let\sub\blacktriangle 
\renewcommand{\blacktriangle}{\textcolor{grassgreen}{\sub}}
\let\sup\bullet

\renewcommand{\bullet}{\textcolor{darkblue}{\sup}}

\newcommand{\axis}[1]{\mathbf{\hat #1}}

\begin{document}
\title{Analytic Pulse Sequence Construction for Exchange-Only Quantum Computation}

\author{Daniel Zeuch,$^{1,2}$ R. Cipri,$^{1}$ and N.E. Bonesteel$^{1}$}

\affiliation{$^1$Department of Physics and National High Magnetic
Field Laboratory, Florida State University, Tallahassee, Florida 32310,
USA\\
$^2$Department of Physics, University of Konstanz, D-78457 Konstanz, Germany
}


\begin{abstract}
We present pulse sequences for two-qubit gates acting on encoded qubits for exchange-only quantum computation. Previous work finding such sequences has always required numerical methods due to the large search space of unitary operators acting on the space of the encoded qubits.  By contrast,  our construction can be understood entirely in terms of three-dimensional rotations of effective spin-1/2 pseudospins which allows us to use geometric intuition to determine the required sequence of operations analytically.  The price we pay for this simplification is that, at 39 pulses, our sequences are significantly longer than the best numerically obtained sequences.
\end{abstract}


\maketitle

\section{Introduction}

The ability to adiabatically switch on and off, or ``pulse," the isotropic exchange interaction, $J{\bf S}_1 \cdot {\bf S}_2$, between pairs of spin-1/2 particles is a promising resource for quantum computation.\cite{loss98}  Such pulsed exchange has been demonstrated experimentally for electron spins in double quantum dots,\cite{petta05} as well as cold atoms trapped in optical lattices.\cite{anderlini07}  The exchange interaction is purely isotropic and so cannot change the total spin of the system it acts on and thus cannot be used to carry out arbitrary unitary operations.  Nevertheless, the ability to pulse the exchange interaction coherently is a sufficient resource for universal quantum computation, provided the logical qubits of the computer are suitably encoded.\cite{bacon00,kempe01}

DiVincenzo et al.\cite{divincenzo00} presented the first explicit scheme for carrying out universal quantum computation using only pulsed exchange.  In this scheme, each qubit is encoded into the two-dimensional Hilbert space of three spin-1/2 particles with total spin fixed to be $1/2$ and polarized along a given direction. For a linear array of spin-1/2 particles, arbitrary single-qubit rotations can then be carried out by performing a sequence of up to four exchange pulses between nearest-neighbor spins within a given encoded qubit.  

There has been remarkable experimental progress on the implementation of such three-spin qubits using electron spins in triple quantum dots.\cite{laird10,takakura10,gaudreau12,hsieh12,medford13_nn}  A related scheme, based on the so-called resonant exchange qubit,\cite{taylor13} in which the exchange interactions between spins within the qubit are kept ``always on," has also recently been demonstrated.\cite{medford13_prl} These resonant exchange qubits offer resistance to leakage out of the encoded qubit space and the possibility for carrying out two-qubit gates with a single exchange pulse.\cite{doherty13}  In the present work, as in Ref.~\onlinecite{divincenzo00}, we assume the exchange interaction between spins is zero except when pulsing.  In this case two-qubit gates require nontrivial sequences of many exchange pulses to avoid leakage out of the encoded space.

By performing a numerical search, DiVincenzo et al.\cite{divincenzo00} were able to find a sequence of 19 nearest-neighbor exchange pulses for a linear array of spins which carries out a two-qubit gate locally equivalent to a controlled-NOT (CNOT) gate (i.e., a CNOT gate up to single-qubit rotations) on two three-spin qubits.  This numerically obtained sequence was later confirmed to be exact.\cite{kawano05} The set of single-qubit rotations and CNOT gates is a standard universal gate set for quantum computation, and so these pulse sequences can be used to perform any quantum algorithm.\cite{nielsen10}

A key requirement in the CNOT construction of Ref.~\onlinecite{divincenzo00} is that the total spin of all six spin-1/2 particles forming the two encoded qubits acted on by the gate must be 1. As pointed out in the same reference, for electron spins this condition can be forced by initializing the qubits in an external magnetic field.  This total spin requirement cannot be relaxed, because if the total spin of all six particles is 0 then the 19-pulse sequence does not result in the same two-qubit gate and, in fact, leads to leakage out of the encoded qubit space. 

More recently, Fong and Wandzura\cite{fong11} found a sequence of nearest-neighbor exchange pulses, again for a linear array of spins, which performs the same two-qubit gate (also locally equivalent to CNOT) in both the total spin 0 and total spin 1 sectors.  Remarkably, with 18 pulses, this sequence is shorter than the 19-pulse sequence of Ref.~\onlinecite{divincenzo00}.  Although this sequence was obtained by numerical minimization of a cost function using a genetic algorithm, the final result is exact and has a particularly elegant form consisting of $\sqrt{{\rm SWAP}}$, inverse $\sqrt{{\rm SWAP}}$, and SWAP pulses.  Related two-qubit gate sequences with fewer pulses (16 and 14) have since been found for geometries other than linear arrays of spins.\cite{setiawan13}

In this paper we construct a family of sequences consisting of 39 nearest-neighbor exchange pulses on a linear array of spins which perform entangling two-qubit gates on three-spin qubits, including a gate which is locally equivalent to CNOT.  The main new feature of our construction is that it can be carried out purely analytically, requiring at most the solution of a transcendental equation in one variable.  Unlike the 19-pulse sequence of Ref.~\onlinecite{divincenzo00}, but like the 18-pulse sequence of Fong and Wandzura,\cite{fong11} the action of our 39-pulse sequences are independent of the total spin of the two encoded qubits.  Indeed, we point out that any pulse sequence which carries out a leakage-free two-qubit gate in the total spin-1 sector while acting on only five of the six spins needed to encode the qubits (which is the case for our sequences, as well as those found by Fong and Wandzura\cite{fong11} and in Ref.~\onlinecite{setiawan13}, but not for the sequence of Ref.~\onlinecite{divincenzo00} which acts on all six spins) will perform the same two-qubit gate in the total spin-0 sector.  Using such sequences eliminates the need to initialize encoded qubits in a magnetic field.

\begin{figure}
	\includegraphics[width=\columnwidth]{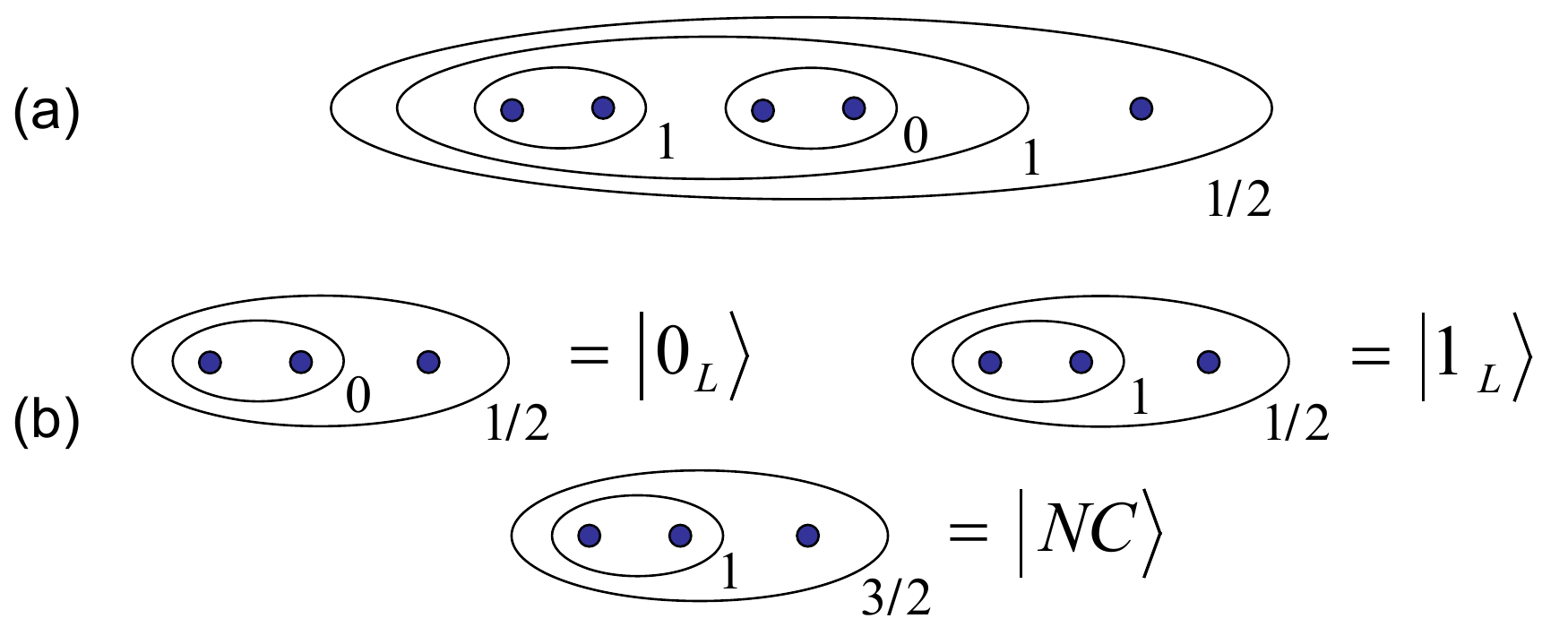}%
	\caption{(color online) (a) Example state of five spin-1/2 particles, where each $\bullet$ represents one particle and the number next to each oval gives the total enclosed spin.  (b) Qubit encoding using three spin-1/2 particles.  States with total spin 3/2 are noncomputational states.}
	\label{basis}%
\end{figure}

\section{Hilbert Space and Qubit Encoding}
\label{hilbert}

Because the isotropic exchange interaction between pairs of spin-1/2 particles is rotationally invariant, any unitary operation carried out purely by pulsing this interaction can be described entirely in terms of total spin quantum numbers, with no reference to $S_z$ quantum numbers.

Figure \ref{basis}(a) illustrates a notation which exploits this fact.  This notation is inspired by that used in Refs.~\onlinecite{bonesteel05,hormozi07} for non-Abelian anyons when finding braiding patterns for topological quantum computation, a problem closely related to that of finding pulse sequences for exchange-only quantum computation.  Here, spin-1/2 particles are represented by solid dots enclosed in ovals labeled by the total spin of the enclosed particles.  Any choice of non-intersecting ovals for which each oval encloses two particles, two ovals, or one of each, amounts to a basis choice.  The basis states correspond to all possible labelings of ovals consistent with the triangle rule for adding spin quantum numbers.   When referring to these basis states in the text we will use parentheses to represent ovals so, e.g., the state shown in Fig.~\ref{basis}(a) would be written $(((\bullet\,\bullet)_1(\bullet\,\bullet)_0)_1\bullet)_{1/2}$ where the symbol $\bullet$ denotes a spin-1/2 particle.  It is always possible to change bases from one set of ovals to another by using the appropriate spin recoupling coefficients.\cite{sign_convention}  

A multi-spin state with total spin $S$ (i.e. the label of the oval enclosing all the particles is $S$) has a $(2S+1)$-fold degeneracy associated with the possible values of the $S_z$-component.  However, as emphasized above, all spin operations we consider for exchange-only quantum computation are rotationally invariant, so at no point will it be necessary to refer to these $S_z$ quantum numbers.  In what follows we will therefore treat states like $\bullet$ or $(\bullet\,\bullet)_1$ as single states in Hilbert space, even though when the $S_z$ degeneracy is counted they are twofold and threefold degenerate, respectively. 

To carry out exchange-only quantum computation it is necessary to use suitably encoded logical qubits.\cite{bacon00}  The basis states for the three-spin qubit encoding of Ref.~\onlinecite{divincenzo00} are shown in Fig.~\ref{basis}(b).  In this encoding, the logical qubit states are those with total spin 1/2, with the logical $|0_L\rangle$ and logical $|1_L\rangle$ corresponding, respectively, to the states for which two of the particles are in a singlet or a triplet. The choice of the two particles whose total spin determines the state of the logical qubit is, of course, purely a basis choice. The price one pays for this qubit encoding is that there is a noncomputational state, denoted $|NC\rangle$ in Fig.~\ref{basis}(b), in which the total spin of the three particles is 3/2.  

Transitions from the computational space to the noncomputational space are known as leakage errors.  When carrying out single-qubit rotations by pulsing the exchange interaction within a given encoded qubit, the total spin of that qubit is unchanged and there are no leakage errors.  However, carrying out two-qubits gates requires some pulses that act on spins from each qubit.  Such pulses alter the total spin of each encoded qubit and thus induce transitions into the noncomputational space.  It is therefore a nontrivial problem to determine pulse sequences which carry out leakage-free entangling two-qubit gates.  

\begin{figure}%
	\includegraphics[width=\columnwidth]{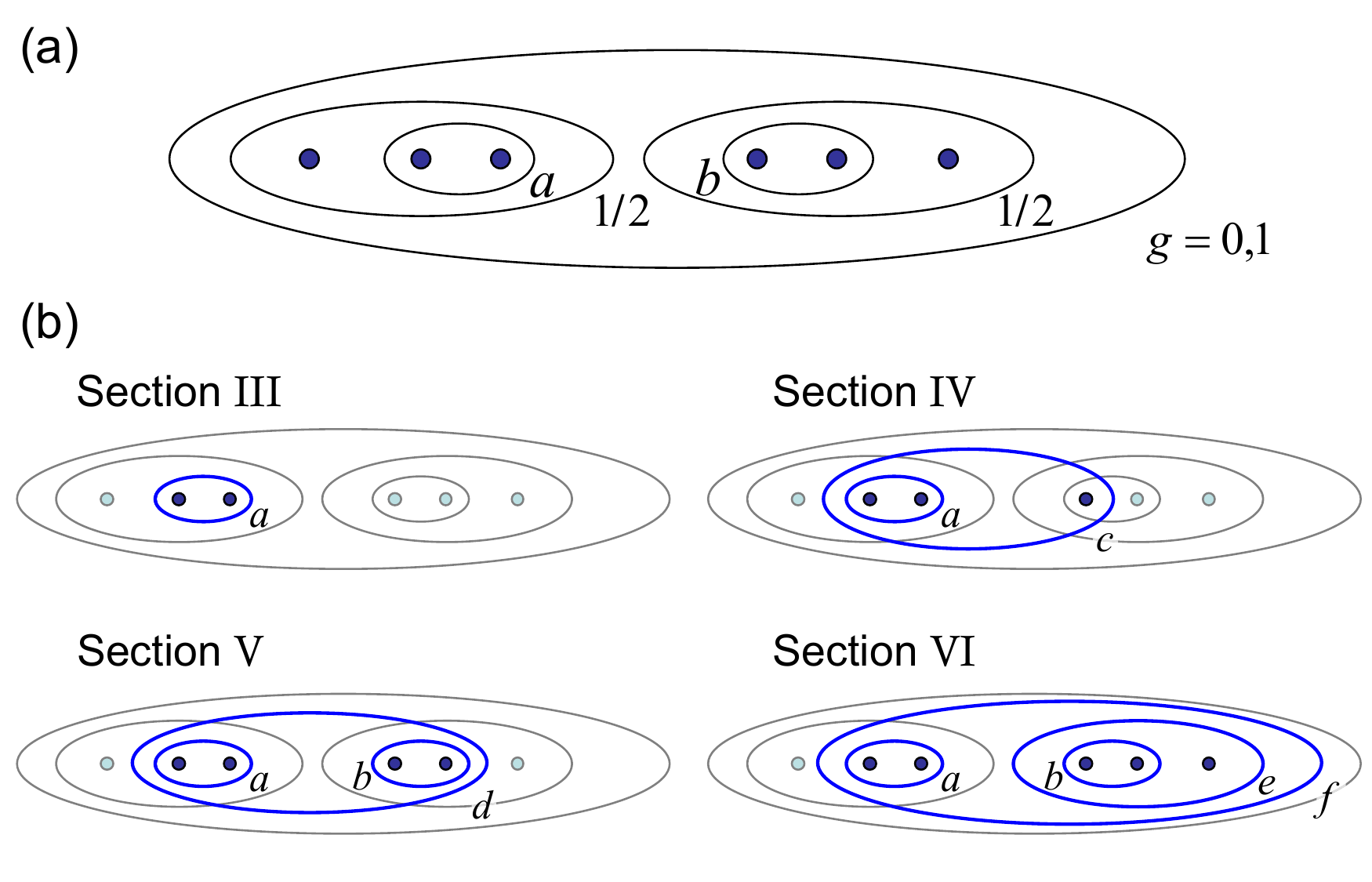}%
	\caption{(color online) (a) Two three-spin qubits in states $a$ and $b$ with total spin $g=0$ or 1.  (b) Relevant spins (circled in blue) and spin quantum numbers referred to in Sec.~\ref{2spins} through Sec.~\ref{5spins}.}%
	\label{twoqubits}%
\end{figure}

Figure \ref{twoqubits}(a) shows two logical qubits each of which has total spin 1/2, so the total spin of all six spin-1/2 particles, labeled $g$, can be either 0 or 1.  In our construction we assume these spins form a linear array and only consider nearest-neighbor exchange pulses.  The choice of qubit bases in the figure is convenient for our two-qubit gate construction.  The full Hilbert spaces of the $g=0$ and $g=1$ sectors are five- and nine-dimensional, respectively, where, as described above, we ignore the $S_z$ degeneracy.  The set of unitary operators acting on this space is then $SU(5) \oplus SU(9)$, once irrelevant overall phase factors are removed.  The number of independent parameters appearing in these unitary operators are $24 = 5^2 -1$ (for $g=0$) and $80 = 9^2 -1$ (for $g=1$).  It is because of the enormous size of these high-dimensional search spaces that all previous work finding pulse sequences for two-qubit gates has been numerical, even when the result has the elegant form of the Fong-Wandzura sequence. 

An outline of our analytic approach to constructing pulse sequences is illustrated in Fig.~\ref{twoqubits}(b).  After establishing the fundamental resource---the exchange interaction between two spin-1/2 particles---we consider the Hilbert spaces of three spins, four spins, and finally five spins.  At each stage of our construction we work with a restricted set of operations which allows us to work entirely in effective Hilbert spaces which are at most two-dimensional, i.e. that of a spin-1/2 pseudospin.  The space of operations is then that of simple three-dimensional rotations and this allows us to use geometric intuition to analytically determine the required pulse sequences.  

Our construction results in a controlled-phase (CPhase) gate which is diagonal in the $ab$ basis for the two qubits shown in Fig.~\ref{twoqubits}(a) and which applies a phase factor of $e^{-i\phi}$ to the state with $ab=11$ while multiplying the states $ab = 00,01,$ and 10 by 1.  We are able to set $\phi$ to any desired phase and the case $\phi = \pi$ yields a gate which is locally equivalent to CNOT.  Two examples of the resulting pulse sequences, which consist of 39 pulses and either one or two single-qubit rotation pulses, are given in Sec.~\ref{full}. 

\begin{figure}[t]%
	\includegraphics[width=\columnwidth]{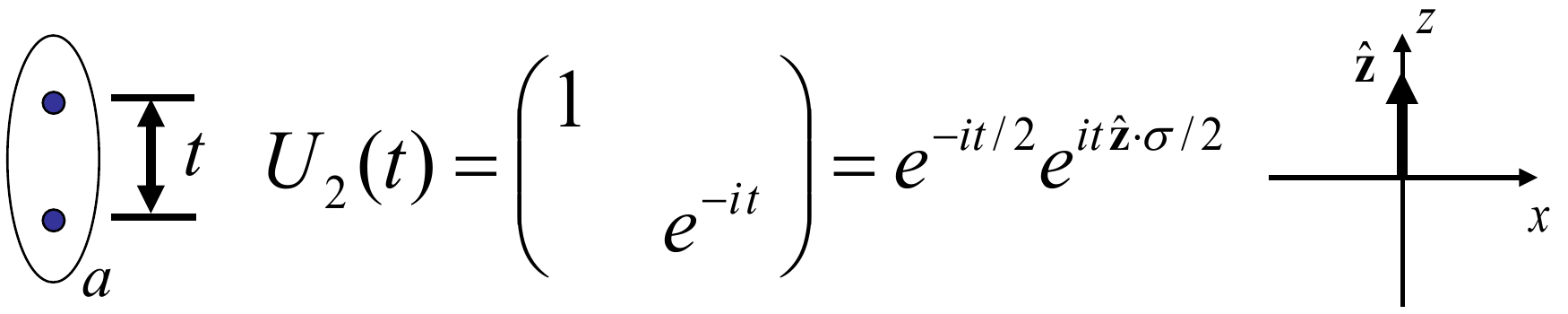}%
	\caption{(color online) Exchange pulse between two spin-1/2 particles, represented by a double arrow labeled by the pulse duration $t$ defined in the text, which produces the operation $U_2(t)$.  In the basis $a=\{0,1\}$, the matrix representation of $U_2(t)$ is a $z$-axis rotation in pseudospin space with $\uparrow  = (\bullet\,\bullet)_0$ and $\downarrow = (\bullet\,\bullet)_1$.}%
	\label{U2}%
\end{figure}

\section{Two Spins}
\label{2spins}

We begin our construction by considering an exchange pulse between two nearest-neighbor spins (e.g., the spins circled in Fig.~\ref{twoqubits}(b) in the diagram labeled ``Section \ref{2spins}").  The effect of such a pulse generated by applying the Hamiltonian $J {\bf S}_1 \cdot {\bf S}_2$ for duration $t$, measured in units of $1/J$ ($\hbar = 1$), is illustrated in Fig.~\ref{U2}.\cite{integrated_pulse}  The matrix representation of the resulting unitary operation in the $(\bullet\,\bullet)_a$ basis with $a=\{0,1\}$, i.e. the singlet-triplet basis where, as described in the previous section, we ignore the $S_z$ degeneracy, is
\begin{eqnarray}
	U_2(t) &=& e^{-i t\left({\bf S}_1 \cdot {\bf S}_2 + \frac34\right)} \nonumber \\
	&=&
	\left(\begin{array}{cc}
		1\\
		&	e^{-it}
	\end{array}\right) 
= e^{-it/2} e^{it \axis z \cdot \boldsymbol{\sigma}/2}.
	\label{eq:U2}
\end{eqnarray}
Here $\boldsymbol{\sigma} = (\sigma_x,\sigma_y,\sigma_z)$ is the Pauli vector and the additive constant 3/4 in the exponent gives a convenient choice for the irrelevant overall phase factor.  If we view the states $(\bullet\,\bullet)_0$ and $(\bullet\,\bullet)_1$ as the $\uparrow$ and $\downarrow$ states, respectively, of a pseudospin then this operation is a $z$-axis rotation in pseudospin space through the angle $t$ multiplied by a phase factor.  
 
Our convention throughout will be that positive pseudospin rotation angles correspond to left-handed rotations about the given axis (i.e., a rotation through angle $t$ about an axis $\axis n$ corresponds to the $SU(2)$ operation $U = e^{it\axis n \cdot {\boldsymbol{\sigma}}/2}$).  The duration of each pulse is positive and can always be taken to be in the range $0 <  t < 2\pi$.  For the inverse of an exchange pulse of duration $t$ we pulse for duration $s = 2\pi - t$.

\begin{figure}
	\begin{center}
	\includegraphics[width=\columnwidth]{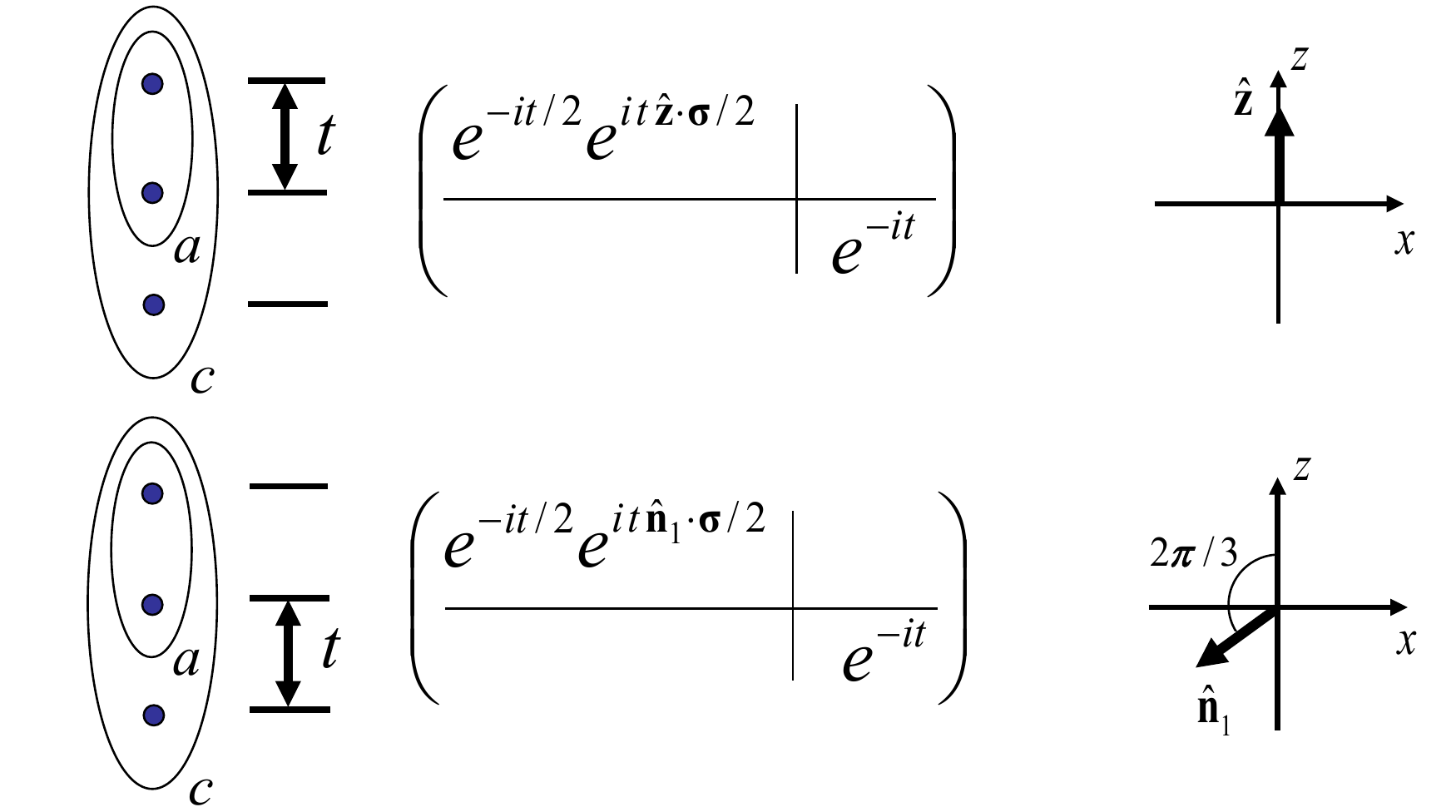}
	\caption{(color online) Nearest-neighbor exchange pulses (denoted $U_2$ in the text) and their matrix representations in the $ac=\{0\frac12,1\frac12 | 1\frac32\}$ basis.  Each $3\times 3$ matrix is block-diagonal, consisting of a $2\times 2$ sector with $c=1/2$ and a one-dimensional sector with $c=3/2$.  In the $c=1/2$ sector the pulses produce rotations about either $\axis z$ or $\axis n_1$ for a pseudospin where $\uparrow = ((\bullet\,\bullet)_0\,\bullet)_{1/2}$ and $\downarrow = ((\bullet\,\bullet)_1\,\bullet)_{1/2}$.} 
	\label{resources3}
  \end{center}
\end{figure}

\section{Three Spins}
\label{3spins}

Figure \ref{resources3} shows the action of two different nearest-neighbor exchange pulses on the Hilbert space of three spin-1/2 particles (e.g., the three spins circled in Fig.~\ref{twoqubits}(b) in the diagram labeled ``Section \ref{3spins}").  As described in Sec.~\ref{hilbert}, the choice of labeled ovals corresponds to a particular basis choice.  The three-spin basis shown in Fig.~\ref{resources3} consists of the states $((\bullet\,\bullet)_a\bullet)_c$ where $ac = 0\frac12,1\frac12,$ and $1\frac32$.   For clarity, when referring to vertically aligned spins in a given figure the convention is that topmost in the figure corresponds to leftmost in the text.

Matrix representations of the unitary operations produced by the exchange pulses are also shown in Fig.~\ref{resources3}.  These matrices are expressed in the $((\bullet\,\bullet)_a\bullet)_c$ basis with $ac = \{0\frac{1}{2},1\frac{1}{2}|1\frac{3}{2}\}$ and consist of a $2\times 2$ block acting on the total spin $c=1/2$ sector and a phase factor multiplying the $c=3/2$ state.  

We describe the two-dimensional $c=1/2$ sector in terms of a pseudospin with $\uparrow = ((\bullet\,\bullet)_0\bullet)_{1/2}$ and $\downarrow = ((\bullet\,\bullet)_1\bullet)_{1/2}$. The unitary operations shown in Fig.~\ref{resources3} are then pseudospin rotations about two different axes.  Pulsing the top two spins results in an operation that is diagonal in $a$ and hence is a rotation about the $z$-axis. In the $((\bullet\,\bullet)_a\bullet)_{1/2}$ basis with $a = \{0, 1\}$, the matrix representation of this operation is the same as that given in \eqref{eq:U2}.    

\begin{figure}[t]
	\begin{center}
	\includegraphics[width=\columnwidth]{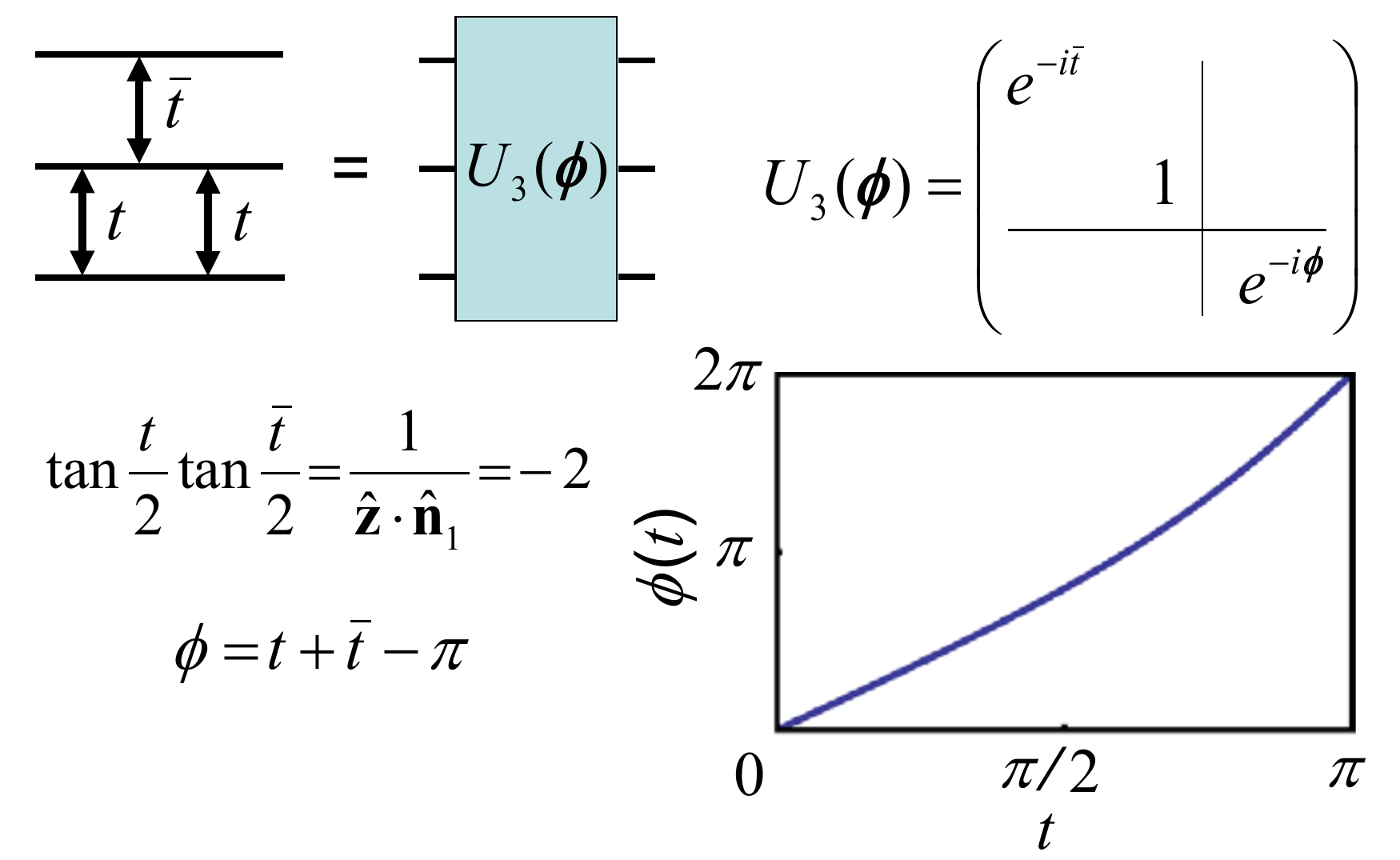}
	\caption{(color online) Sequence of three exchange pulses for $U_3(\phi)$, a diagonal operation in the $((\bullet\,\bullet)_a\bullet)_c$ basis, shown for $ac=\{0\frac12,1\frac12|1\frac32\}$.  $U_3(\phi)$ introduces a phase difference $\phi$ between the states $((\bullet\,\bullet)_1\bullet)_{1/2}$ and $((\bullet\,\bullet)_1\bullet)_{3/2}$.    The graph of $\phi$ vs.\ $t$ shows that an arbitrary phase $\phi$ can be generated by choosing $t$ appropriately.}
	\label{U3}
	\end{center}
\end{figure}

Likewise, the matrix representation of an exchange pulse between the bottom two spins (see Fig.~\ref{resources3}) in the $(\bullet(\bullet\,\bullet)_{a^\prime})_{1/2}$ basis with $a^\prime = \{0,1\}$ is
\begin{eqnarray}
{U}_{2,a^\prime}^{c=1/2}(t) = \left(\begin{array}{cc} 1 & \\ & e^{-it}\end{array}\right) = e^{-it/2} e^{i t \axis z \cdot \boldsymbol{\sigma}/2}.
\end{eqnarray}
Here the notation ${U}_{2,a^\prime}^{c=1/2}$ indicates the matrix representation of $U_2$ (in this case the unitary operation produced by pulsing the exchange interaction between the bottom two spins) in the $a^\prime$ basis in the sector with total spin $c=1/2$.  To find the matrix in the original $((\bullet\,\bullet)_a\bullet)_{1/2}$ basis we perform the basis change
\begin{eqnarray}
((\bullet\,\bullet)_a\bullet)_{1/2} = \sum_{a^\prime} F_{1,aa^\prime}(\bullet(\bullet\,\bullet)_{a^\prime})_{1/2},
\end{eqnarray}
where the matrix elements
\begin{eqnarray}
F_{1,aa^\prime} = \langle (\bullet(\bullet\,\bullet)_{a^\prime})_{1/2} | ((\bullet\,\bullet)_a\,\bullet)_{1/2} \rangle
\end{eqnarray}
are recoupling coefficients for three spin-1/2 particles with total spin 1/2.  $F_{1,aa^\prime}$ can be expressed as a $2\times 2$ matrix which transforms from the $a^\prime = \{0,1\}$ basis to the $a = \{0,1\}$ basis,
\begin{eqnarray}
F_{1} =
\left(\begin{array}{cc} -1/2 & \sqrt{3}/2 \\ \sqrt{3}/2 & 1/2\end{array}\right) = \hat {\bf f}_{1} \cdot \boldsymbol{\sigma},
\end{eqnarray}
where $\hat{\bf f}_{1} = (\sqrt{3}/2,0,-1/2)$.  The action of pulsing the exchange interaction between the bottom two spins in the $((\bullet\,\bullet)_a\bullet)_{1/2}$ basis with $a = \{0,1\}$ is then
\begin{eqnarray}
U_{2,a}^{c=1/2}(t) = F_1 U_{2,a^\prime}^{c=1/2}(t) F_{1} = e^{-it/2} e^{it{\axis n_1} \cdot \boldsymbol{\sigma}/2},
\end{eqnarray}
where $F_1 = {F_1}^\dagger$.  The rotation axis $\axis n_1 = 2 \axis f_1 (\axis f_1 \cdot \axis z) - \axis z$ makes an angle $\cos^{-1} \axis n_1 \cdot \axis z = -\frac{2\pi}3$ with the $z$-axis, as shown in Fig.~\ref{resources3}.

The $c=3/2$ sector consists of a single state which can be expressed equivalently either as $((\bullet\,\bullet)_1\bullet)_{3/2}$ or $(\bullet(\bullet\,\bullet)_1)_{3/2}$.  Consulting \eqref{U2} for the case $a=1$ we see that both exchange pulses of duration $t$ shown in Fig.~\ref{resources3} multiply this state by a phase factor of $e^{-it}$.  Thus the $ac = 1\frac32$ diagonal element of the corresponding matrix representations is $e^{-it}$.

\begin{figure}
	\includegraphics[width=\columnwidth]{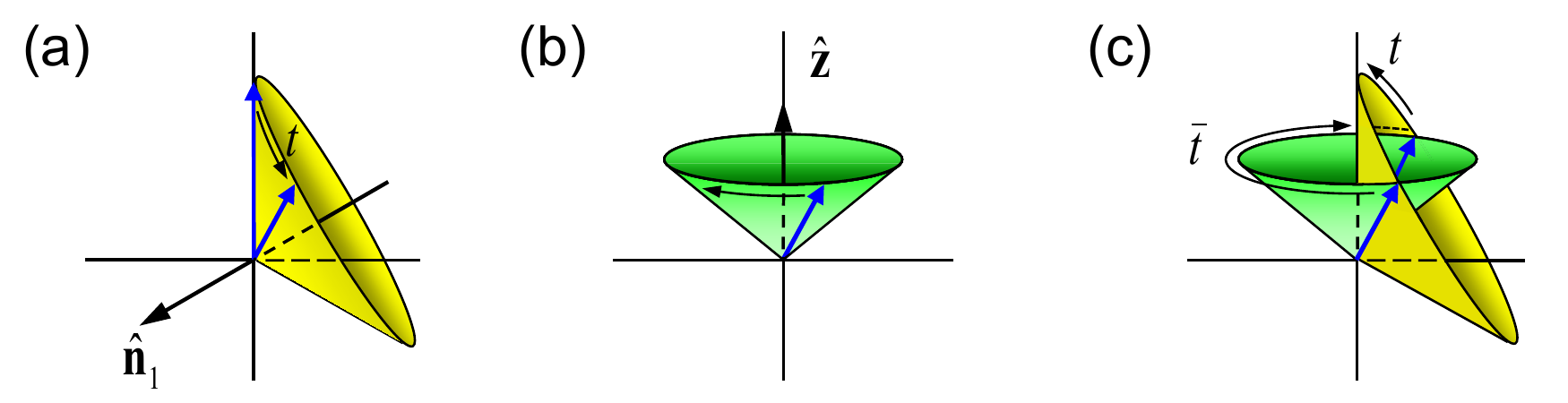}%
	\caption{(color online) Actions of the three rotations in the $c=1/2$ pseudospin sector of the three-pulse sequence for $U_3$ shown in Fig.~\ref{U3} on the vector $\axis z$. (a) The first pulse of duration $t$ rotates $\axis z$ about the $n_1$-axis to a vector on the yellow cone.  (b) The second pulse of duration $\bar t$ rotates the resulting vector about the $z$-axis on the green cone.  (c) Provided $\bar t$ is chosen (by solving \eqref{ttbar}) so that after the first two rotations the resulting vector is on the intersection of the green and yellow cones, the third pulse of duration $t$ will rotate the vector about the $n_1$-axis back to $\axis z$.  Because $\axis z$ is unchanged by this sequence, the resulting rotation is about the $z$-axis.}%
	\label{cone}%
\end{figure}

Figure \ref{U3} shows a key three-pulse sequence used throughout our construction. The resulting unitary operation is denoted $U_3$.  This pulse sequence is designed so that the matrix representation of $U_3$ is diagonal in the $((\bullet\,\bullet)_a\bullet)_c$ basis, as shown in Fig.~\ref{U3} (up to an irrelevant overall phase factor, chosen so that the state $((\bullet\,\bullet)_1\bullet)_{1/2}$ is multiplied by 1).  This allows us to treat the Hilbert space with $a=0$ and $a=1$ separately,  while at the same time generating a phase difference of $\phi$ between the states $((\bullet\,\bullet)_1\bullet)_{1/2}$ and $((\bullet\,\bullet)_1\bullet)_{3/2}$. This phase difference is central to our construction and in what follows we will often write $U_3$ as a function of this phase, $U_3(\phi)$.

In the $c=1/2$ sector, the pulse sequence for $U_3$ carries out three pseudospin rotations about first the $n_1$-, then $z$-, and again the $n_1$-axis through angles $t$, $\bar t$, and $t$, respectively.  This sequence is chosen so that it results in a net rotation about the $z$-axis, and hence is diagonal in the $((\bullet\,\bullet)_a\bullet)_{1/2}$ basis.  To find the relation between $t$ and $\bar t$ we determine the condition under which the vector $\axis z$ is unchanged under these three rotations.  The yellow cone in Fig.~\ref{cone}(a) shows the set of vectors that $\axis z$ can be transformed into after rotations about the $n_1$-axis by the first pulse.  For a particular choice of the first rotation angle $t$, the green cone in Fig.~\ref{cone}(b) then displays the set of possible outcomes of the second rotation, this time about the $z$-axis.  The third rotation, again about the $n_1$-axis, must bring the transformed vector back to $\axis z$. Figure \ref{cone}(c) shows both that there is only one non-zero choice for the second rotation angle, $\bar t$, and that the final rotation angle must again be $t$.  It is a simple geometric exercise to show that the rotation angles $\bar t$ and $t$ are related by
\begin{eqnarray}
	\tan \frac{t}{2} \tan \frac{\bar t}{2} = \frac{1}{{\axis z} \cdot {\axis n_1}} = -2.\label{ttbar}
\end{eqnarray}
Furthermore, Fig.~\ref{cone} clarifies that the $t,\bar t,t$ sequences are the only nontrivial sequences of three rotations that result in an effective $z$-axis rotation.

The sequence $t, \bar t, t$ produces the phase difference
\begin{eqnarray}
	\phi = t + \bar t - \pi\label{phit}
\end{eqnarray} 
between the $((\bullet\,\bullet)_1\bullet)_{1/2}$ and $((\bullet\,\bullet)_1\bullet)_{3/2}$ states. As a function of the pulse length $t$, the phase $\phi$ varies monotonically from $0$ to $2\pi$ (see Fig.~\ref{U3}). Thus, to produce $U_3(\phi)$ for a desired $\phi$ one need only solve for $t$ and $\bar t$ using \eqref{ttbar} and \eqref{phit}. For a given $\phi$ there are two solutions, one with $0 \le t < \pi \le \bar t < 2\pi$, and another with $t \leftrightarrow \bar t$ so that $0 \le \bar t < \pi \le t < 2\pi$.  The total duration of the $t,\bar t, t$ sequence with $t < \bar t$ is shorter than the sequence with $t > \bar t$, and we refer to the former as the short sequence and the latter as the long sequence.  The only difference between the $U_3(\phi)$ operations produced by the short sequence and long sequence is the value of the phase factor $e^{-i\bar t}$ applied to the single state with $a=0$, $((\bullet\,\bullet)_0\bullet)_{1/2}$.  In our two-qubit gate construction we will see that the only effect the choice of this phase factor has is to determine the single-qubit rotations needed to bring the final gate to an exact CPhase form.  We are thus free to use either the short or long sequence for each $U_3$ that appears in our construction (see Sec.~\ref{full}).\cite{ttbart} 

\section{Four Spins}
\label{4spins}

In this section we turn to the four spins highlighted in Fig.~\ref{twoqubits}(b) (labeled ``Section V"), $((\bullet\,\bullet)_a(\bullet\,\bullet)_b)_d$ where $a$ and $b$ determine the states of the two logical qubits shown in Fig.~\ref{twoqubits}(a).

\begin{figure}
	\begin{center}
	\includegraphics[width=\columnwidth]{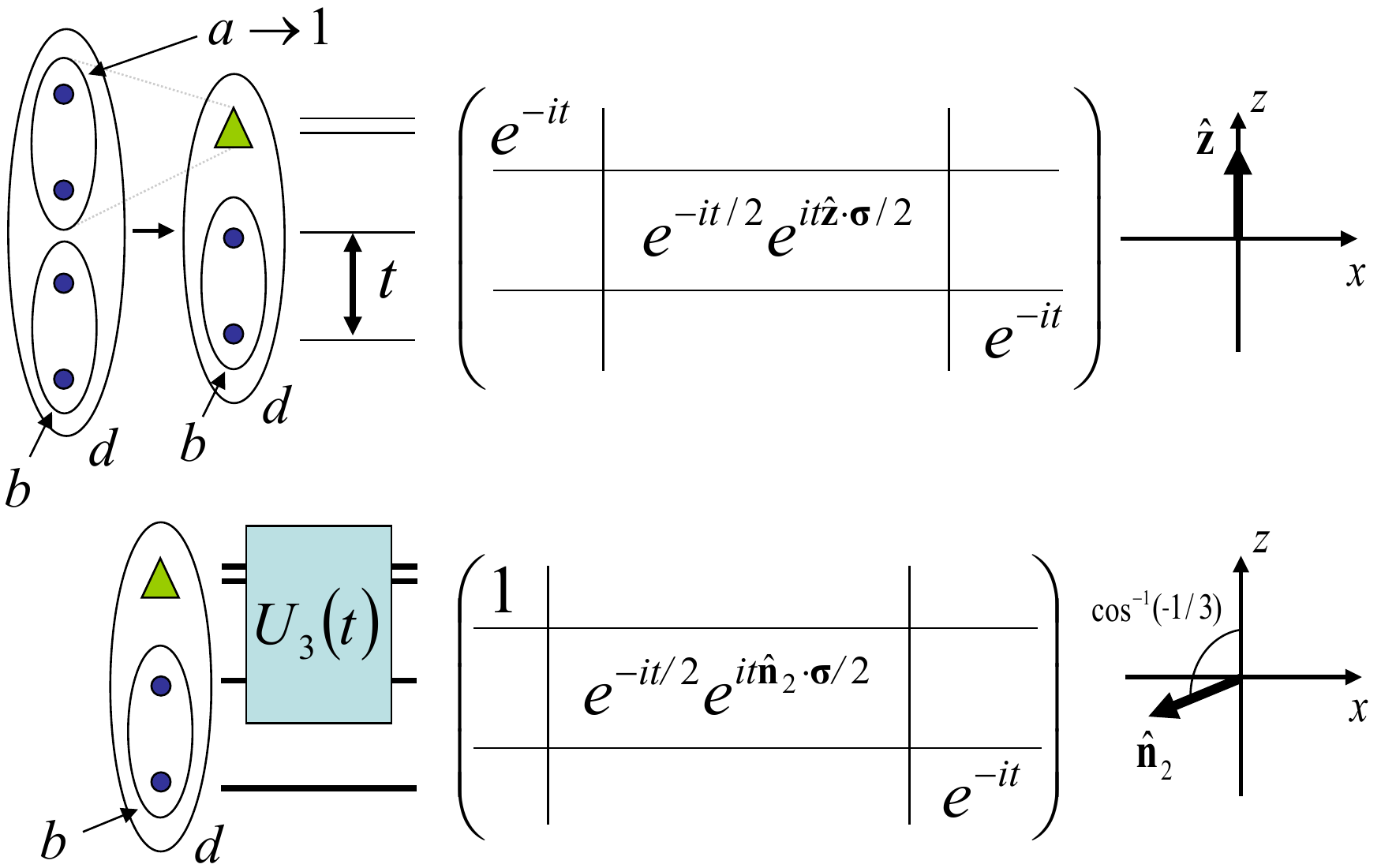}
	\caption{(color online) Two operations, a simple exchange pulse [$U_2(t)$] and $U_3(t)$, acting on the Hilbert space of four spins.  Both operations conserve $a$ and act trivially on states with $a=0$.  This allows us to focus on the case $a=1$ by replacing the two spins with total spin $a$ by an effective spin-1 particle represented by $\blacktriangle$.  The matrix representations of $U_2(t)$ and $U_3(t)$ are then given in the $bd=\{10|01,11|12\}$ basis.  In the $d=1$ sector, $U_2(t)$ and $U_3(t)$ carry out rotations about ${\axis z}$ and ${\axis n_2}$, respectively, for a pseudospin where $\uparrow = (\blacktriangle(\bullet\,\bullet)_0)_1 $ and $\downarrow = (\blacktriangle(\bullet\,\bullet)_1)_1$.}
	\label{resources4}
	\end{center}
\end{figure}

The full Hilbert space of four spin-1/2 particles [as usual, not counting the $S_z$ degeneracy] is six-dimensional with one two-dimensional sector (total spin 0), one three-dimensional sector (total spin 1) and one one-dimensional sector (total spin 2).  We reduce the nontrivial Hilbert space to that of a single spin-1/2 pseudospin by restricting ourselves to the use of the two operations shown in Fig.~\ref{resources4}.  One operation is the $U_3$ sequence described in Sec.~\ref{3spins} acting on the top three spins, the other is a simple exchange pulse $U_2$ between the bottom two spins. 

Throughout our entire two-qubit gate construction (excluding single-qubit rotations), the top two spins with total spin labeled $a$, referring to Fig.~\ref{resources4}, will {\it only} be acted on by $U_3$ operations.  Because this operation is diagonal in the $((\bullet\,\bullet)_a\bullet)_c$ basis, the value of $a$ is conserved and we can treat the cases $a=0$ and $a=1$ separately.  For the case $a=0$ the top three spins are always in the state $((\bullet\,\bullet)_0\bullet)_{1/2}$.  It follows that $U_3$ acts as the identity times a phase factor on {\it all} states with $a=0$ in the full Hilbert space of the two encoded qubits.  Provided we keep track of this $a=0$ phase (which will depend on whether we use the long or short sequence for $U_3$) as it accumulates we are free to focus on the case $a=1$ for which $U_3$ acts nontrivially on a two-dimensional Hilbert space.  At the end of our construction the $a=0$ phase factor can always be set to 1 by a single-qubit rotation acting on the left qubit in Fig.~\ref{twoqubits}(a).  

Since we need only consider the case $a=1$ in what follows we can represent the top two spins as a single spin-1 particle, as shown in Fig.~\ref{resources4}.  The basis states can then be written
\begin{eqnarray}
	((\bullet\,\bullet)_{a=1}(\bullet\,\bullet)_b)_d \rightarrow (\blacktriangle(\bullet\,\bullet)_b)_d,
	\label{4basis}
\end{eqnarray}
where the symbol $\blacktriangle$ represents the effective spin-1 particle.  This replacement of two spin-1/2 particles by one spin-1 particle is a key step in our construction.

The $a=1$ Hilbert space---spanned by a spin-1 and two spin-1/2 particles---is four-dimensional, with two one-dimensional sectors (total spin $d=0$ and 2) and one two-dimensional sector (total spin $d=1$).  The effective two-dimensional $d=1$ sector can be viewed in terms of a pseudospin where $\uparrow = (\blacktriangle(\bullet\,\bullet)_{b=0})_1$ and $\downarrow = (\blacktriangle(\bullet\,\bullet)_{b=1})_1$. As shown in Fig.~\ref{resources4}, pulsing the exchange interaction between the bottom two spins for a time $t$ then results in a $z$-axis rotation through angle $t$ of this pseudospin.

The action of $U_3$ on this two-dimensional Hilbert space is first seen most clearly in the $((\blacktriangle\,\bullet)_c\bullet)_{d=1}$ basis with $c=\{\frac12,\frac32$\}.  Consulting Fig.~\ref{U3} for the case $a=1$, we have
\begin{eqnarray}
	U_{3,c}^{d=1}(t) = \left(\begin{array}{cc} 1 & \\ & e^{-it} \end{array} \right) = e^{-it/2} e^{i t \axis z\cdot {\boldsymbol{\sigma}}/2}.
	\label{complaint}
\end{eqnarray}
This operation acts like a nearest-neighbor exchange pulse between our effective spin-1 particle and its neighboring spin-1/2 particle.  However, here the parameter $t$ is not a pulse duration, but rather the value of the phase difference $U_3(t)$ produces between the states $((\blacktriangle\,\bullet)_{1/2}\,\bullet)_1$ and $((\blacktriangle\,\bullet)_{3/2}\,\bullet)_1$, and is best viewed as an ``effective" pulse time.

If we change back to the $(\blacktriangle(\bullet\,\bullet)_b)_{1}$ basis $U_3$ becomes a pseudospin rotation about an axis $\axis n_2$.  To determine $\axis n_2$ we again need to carry out a basis change using the relevant recoupling coefficients, this time for one spin-1 particle and two spin-1/2 particles with total spin 1, 
\begin{eqnarray}
(\blacktriangle(\bullet\,\bullet)_b)_1 = \sum_{c} F_{2,bc} ((\blacktriangle\,\bullet)_c\bullet)_1,
\end{eqnarray}
where $F_{2,bc} = \langle ((\blacktriangle\,\bullet)_c\bullet)_1 | (\blacktriangle(\bullet\,\bullet)_b)_1\rangle$.  The matrix
\begin{eqnarray}
	F_2 = \left(\begin{array}{cc} -1/\sqrt{3} & \sqrt{2/3} \\ \sqrt{2/3} & 1/\sqrt{3}\end{array}\right) = \hat{{\bf f}}_2 \cdot \boldsymbol{\sigma}
	\label{F2}
\end{eqnarray}
then changes bases from $((\blacktriangle\,\bullet)_c\bullet)_1$ with $c=\{\frac12,\frac32\}$ to $(\blacktriangle(\bullet\,\bullet)_b)$ with $b=\{0,1\}$, where $\hat{{\bf f}}_2 = (\sqrt{2/3},0,-1/\sqrt{3})$.   The action of $U_3(t)$ on the $d=1$ sector in the original basis is then
\begin{eqnarray}
U_{3,b}^{d=1}(t) = F_2 U_{3,c}^{d=1}(t) F_2 = e^{-it/2} e^{it{\axis n_2} \cdot \boldsymbol{\sigma}/2},
\end{eqnarray}
where the rotation axis $\axis n_2 = 2 \axis f_2 (\axis f_2 \cdot \axis z) - \axis z$ makes an angle $\cos^{-1} \axis n_2 \cdot \axis z = \cos^{-1} -\frac13$ with the $z$-axis, as shown in Fig.~\ref{resources4}.

\begin{figure}
	\begin{center}
	\includegraphics[width=\columnwidth]{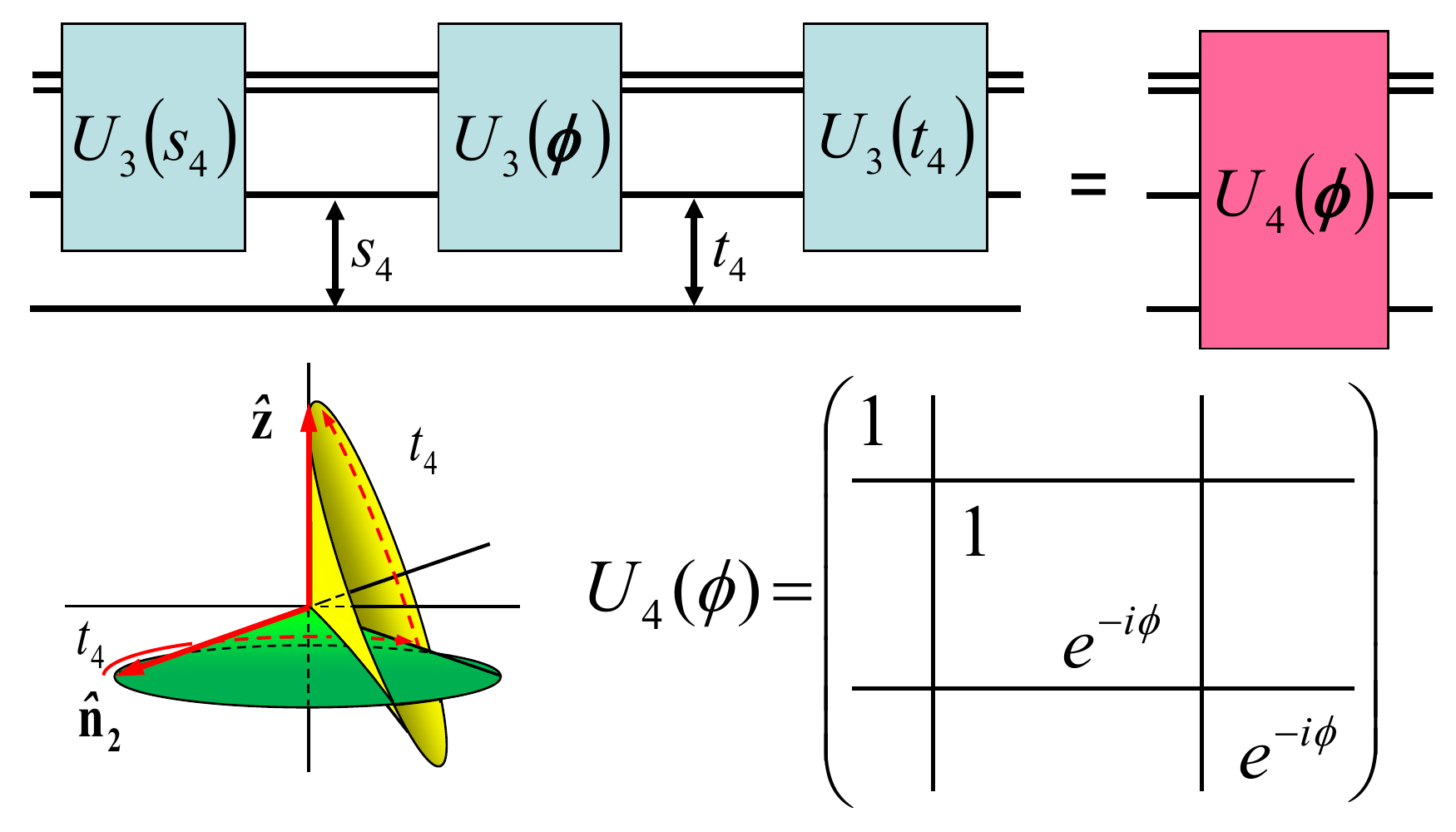}
	\caption{(color online) Sequence of exchange pulses ($U_2$) and $U_3$ operations acting on four spins resulting in the operation $U_4(\phi)$.  The sequence is constructed so that the matrix representation of $U_4(\phi)$ is diagonal in the $(\blacktriangle(\bullet\,\bullet)_b)_d$ basis, shown for $bd = \{10|01,11|12\}$.   The two-step similarity transformation that diagonalizes the $d=1$ block of $U_3(\phi)$ in this basis is illustrated by the two intersecting cones where $t_4=2\pi/3$ and, for the inverse operation, $s_4=2\pi-t_4 = 4\pi/3$.} \label{U4}
	\end{center}
\end{figure}

Finally, note that for the sectors with total spin $d=0$ and 2 the change of bases is trivial: $(\blacktriangle(\bullet\,\bullet)_{b=1})_0 = ((\blacktriangle\,\bullet)_{c=1/2}\,\bullet)_0$ and $(\blacktriangle(\bullet\,\bullet)_{b=1})_2 = ((\blacktriangle\,\bullet)_{c=3/2}\,\bullet)_2$.   Consulting Fig.~\ref{U3}, we see that $U_3(t)$ multiplies the states with $d=0$ and $d=2$ by 1 and $e^{-it}$, respectively, while from \eqref{eq:U2} the exchange pulse $U_2(t)$ acting on the bottom two spins multiplies both states by $e^{-it}$.  The resulting full matrix representations of $U_2(t)$ and $U_3(t)$ are given in Fig.~\ref{resources4}.

At the next level of our construction we will need an operation which is diagonal in the $(\blacktriangle(\bullet\,\bullet)_b)_d$ basis.  This will allow us to treat the Hilbert space with $b=0$ and $b=1$ separately.  The simplest way to produce such a diagonal operation would be to pulse the exchange interaction between the bottom two spins with total spin $b$ (see Fig.~\ref{resources4}).  However, such a pulse will merely correspond to a single-qubit rotation, and therefore is not useful for our two-qubit gate construction.  One way to produce a diagonal operation which is {\it not} equivalent to a single-qubit rotation would be to employ the same three-pulse strategy used in Sec.~\ref{3spins}.  In this case the rotation axis $\axis n_2$ is different (and so the right-hand side of \eqref{ttbar} is $1/\axis n_2 \cdot \axis z = -3$ instead of $1/\axis n_1 \cdot \axis z = -2$), but the geometric argument summarized in Fig.~\ref{cone} still shows that any three-pulse sequence which produces a diagonal matrix must be of the same $t, \bar t, t$ form as $U_3$.  In Appendix \ref{alternate} we show that this construction does indeed produce a diagonal operation in the $(\blacktriangle(\bullet\,\bullet)_b)_d$ basis, but cannot directly be used to produce the required phase differences at the next level of our construction.  Nevertheless, the existence of this three-operation construction does point the way to alternate two-qubit gate constructions, also discussed in Appendix \ref{alternate}.

Given that three operations are not sufficient we turn to sequences with five operations (sequences with four operations are equivalent to sequences with three operations up to a single-qubit rotation).  Figure \ref{U4} shows such a sequence that produces a diagonal operation which will be useful at the next level of our construction.  The sequence has the form $U_4(\phi) = U_3(t_4)U_2(t_4) U_3(\phi) U_2(s_4) U_3(s_4)$, where $s_4 = 2\pi - t_4$ so that $U_2(s_4) = U_2(t_4)^{-1}$ and $U_3(s_4) = U_3(t_4)^{-1}$ in the $a=1$ Hilbert space.  Thus, in this space, $U_4(\phi) = S U_3(\phi) S^{-1}$ where $S = U_3(t_4) U_2(t_4)$.   Written in this way, it is clear that $U_4(\phi)$ is the result of a carrying out a similarity transformation on the $U_3(\phi)$ operation at the center of the sequence.  In the two-dimensional $d=1$ sector this transformation can be understood geometrically as a rotation generated by $U_3(t_4) U_2(t_4)$, two pseudospin rotations about first the $z$-axis and then the $n_2$-axis, both through angle $t_4$.  These rotations act on $\axis n_2$, the rotation axis of $U_3(\phi)$, and are designed to diagonalize $U_3(\phi)$ in the $(\blacktriangle\,(\bullet\,\bullet)_b)_1$ basis by rotating $\axis n_2$ to $\axis z$.

The transformation of the rotation axis of $U_3(\phi)$ from $\axis n_2$ to $\axis z$ is illustrated in Fig.~\ref{U4}.  Rotating $\axis n_2$ ($\axis z$) about the $z$-axis ($n_2$-axis) results in the rotated vector lying somewhere on the green (yellow) cone.  The rotation angle $t_4$ is chosen so that $\axis n_2$ is first rotated about the $z$-axis to where the two cones intersect.  This is then followed by a rotation about the $n_2$-axis through the same angle so that the final rotated vector is $\axis z$.  It is straightforward to calculate the required rotation angle,
\begin{equation}
	t_4 = \cos^{-1} \frac{\axis n_2\cdot \axis z}{\axis n_2\cdot \axis z + 1} = \frac{2\pi}3.
\end{equation}

\begin{figure*}
	\includegraphics[width=2\columnwidth]{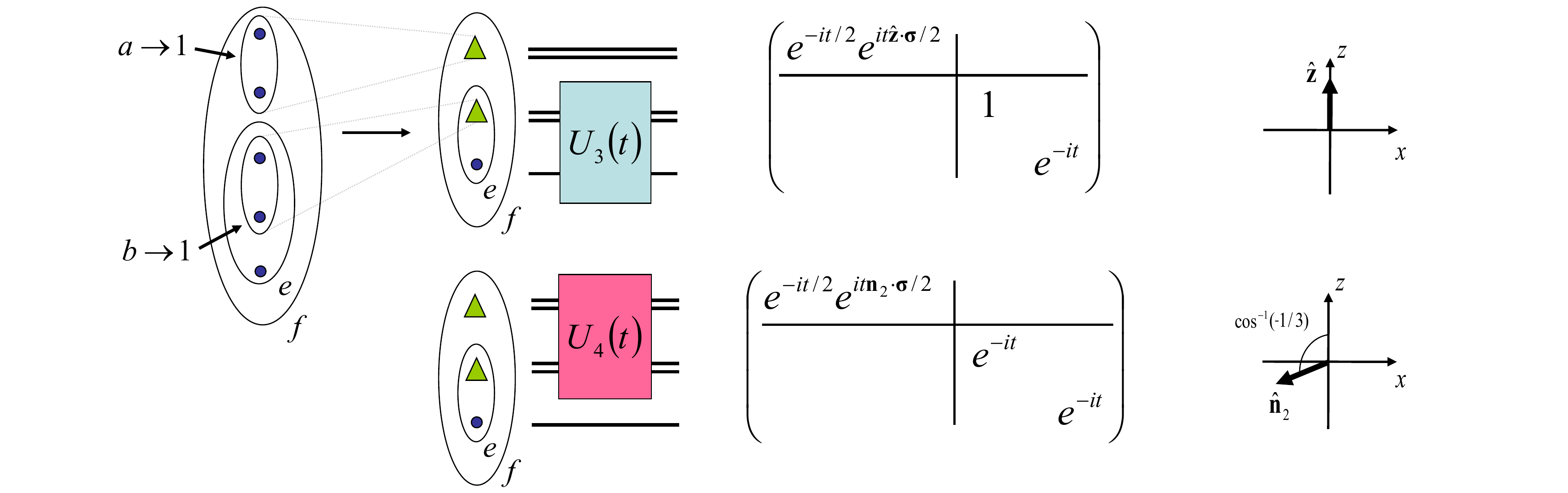}%
	\caption{(color online) Two operations $U_3(t)$ and $U_4(t)$ acting on the Hilbert space of five spins.  Both operations are diagonal in $a$ and $b$ and act trivially on states with $a=0$ and $b=0$.  We are thus able to replace the pairs of spins inside the ovals with total spin $a$ and $b$ by effective spin-1 particles.  The matrix representations of $U_3(t)$ and $U_4(t)$ are then given in the $ef=\{\frac12 \frac12,\frac32 \frac12|\frac12 \frac32,\frac32 \frac32\}$ basis.  In either sector, with $f=1/2$ or 3/2, the operations perform rotations in a pseudospin space with $\uparrow_f = (\blacktriangle(\blacktriangle\,\bullet)_{1/2})_f$ and $\downarrow_f = (\blacktriangle(\blacktriangle\,\bullet)_{3/2})_f$.  The rotation axes in the $f=1/2$ sector are $\axis z$ (for $U_3(t)$) and $\axis n_2$ (for $U_4(t)$), as shown.  In the $f=3/2$ sector, $U_3(t)$ rotates about $\axis z$ and $U_4(t)$ is proportional to the identity.}
	\label{resources5}
\end{figure*}

Due to this similarity transformation $U_4(\phi) = S U_3(\phi) S^{-1}$, the matrix representation of $U_4(\phi)$ in the $d=1$ sector in the $(\blacktriangle(\bullet\,\bullet)_b)_1$ basis with $b=\{0,1\}$ is a $z$-axis rotation,
\begin{eqnarray}
U_{4,b}^{d=1}(\phi) = e^{-i\phi/2} e^{i \phi \axis z \cdot \boldsymbol{\sigma}/2} = 
\left(\begin{array}{cc} 1 & \\ & e^{-i\phi} \end{array}\right).
\label{u4b}
\end{eqnarray}
The full matrix representation of $U_4(\phi)$ in all sectors is shown in Fig.~\ref{U4}.  Since in the one-dimensional sectors with $bd=10$ and $bd=12$ the similarity transformation $U_4(\phi)=S U_3(\phi) S^{-1}$ has no effect on $U_3(\phi)$, the corresponding elements are 1 and $e^{-i\phi}$, respectively (see Fig.~\ref{resources4}).

Let us summarize what we have achieved at this point and what still needs to be done to construct an entangling two-qubit gate. The operation $U_4(\phi)$ multiplies the only $ab=10$ state, $(\blacktriangle(\bullet\,\bullet)_{b=0})_{d=1}$, by 1 while multiplying two of the three $ab=11$ states, $(\blacktriangle(\bullet\,\bullet)_1)_{d=1,\,2}$, by the phase factor $e^{-i\phi}$.  If this operation also multiplied $(\blacktriangle(\bullet\,\bullet)_1)_{d=0}$ by the same phase factor, the action of $U_4(\phi)$ would be to apply a CPhase gate (up to the single-qubit rotation needed to eliminate the $a=0$ phase discussed above) on the two encoded qubits in Fig.~\ref{twoqubits}(a) in which the state $ab=11$ acquires a phase factor $e^{-i\phi}$ while the states $ab=00,01,10$ are multiplied by 1.  However, this is {\it not} the case because $U_4(\phi)$ multiplies $(\blacktriangle(\bullet\,\bullet)_1)_{d=0}$ by 1.  This is consistent with the result of the theorem proved in Appendix \ref{five} which shows that any sequence of exchange pulses acting on only four spins cannot result in a leakage-free entangling two-qubit gate.  To achieve such a gate, we need to consider pulse sequences which act on one more spin.

\section{Five Spins}
\label{5spins}

We now turn to the final stage of our CPhase gate construction which involves five spins.  These spins are highlighted in Fig.~\ref{twoqubits}(b) (labeled ``Section \ref{5spins}") in the $((\bullet\,\bullet)_a((\bullet\,\bullet)_b\,\bullet)_e)_f$ basis with $a$ and $b$ determining the state of the two encoded qubits shown in Fig.~\ref{twoqubits}(a).

The full Hilbert space of five spin-1/2 particles is ten-dimensional and breaks into a five-dimensional sector (total spin 1/2), a four-dimensional sector (total spin 3/2), and a one-dimensional sector (total spin 5/2).  With reference to Fig.~\ref{twoqubits}, note that because the total spin of all six spin-1/2 particles encoding two logical qubits can only be either $g=0$ or $g=1$, the one-dimensional $f=5/2$ sector is not relevant for our two-qubit gate construction.

We use the two operations $U_3$ and $U_4$ shown in Fig.~\ref{resources5} to construct the CPhase gate, where $U_3$ now acts on the bottom three spins and $U_4$ on the top four spins.  In addition to conserving $a$, for the reasons given in Sec.~\ref{4spins}, these operations also conserve $b$.  For the case $b=0$ the top four spins are always in the state $(\blacktriangle(\bullet\,\bullet)_0)_1$ and the bottom three spins are always in the state $((\bullet\,\bullet)_0\bullet)_{1/2}$.  From Fig.~\ref{U4} we then see that $U_4$ acts as the identity and, from the discussion in Sec.~\ref{4spins}, $U_3$ acts as the identity times a phase factor (which depends on whether we use the short or long sequence) on all states with $b=0$ in the full Hilbert space of the two encoded qubits. As for the $a=0$ phase factor discussed in Sec.~\ref{4spins}, if we keep track of this $b=0$ phase factor we are free to focus entirely on the case $b=1$.  The $b=0$ phase factor can then be set to 1 by a single-qubit rotation acting on the qubit on the right in Fig.~\ref{twoqubits}(a).  The only nontrivial case is thus $ab=11$.  To construct a CPhase gate we need to multiply this state with a phase factor of $e^{-i\phi}$.

We exploit the fact that $ab=11$ is the only nontrivial case by working in the reduced Hilbert space of five spin-1/2 particles in which the two spins labeled $a$ and the two spins labeled $b$ are both replaced by effective spin-1 particles, 
\begin{eqnarray}
((\bullet\,\bullet)_{a=1}((\bullet\,\bullet)_{b=1}\,\bullet)_e)_f \rightarrow (\blacktriangle(\blacktriangle\,\bullet)_e)_f,
\end{eqnarray}
as also shown in Fig.~\ref{resources5}.  The effective Hilbert space is then that of one spin-1/2 and two spin-1 particles and has two two-dimensional sectors for $f=1/2$ and $3/2$ (again, as shown above, we need not consider the $f=5/2$ sector).  In both sectors we define a pseudospin $\uparrow_f = (\blacktriangle(\blacktriangle\,\bullet)_{e=1/2})_f$ and $\downarrow_f = (\blacktriangle(\blacktriangle\,\bullet)_{e=3/2})_f$.

The matrix representations of $U_3$ and $U_4$ in the $(\blacktriangle(\blacktriangle\,\bullet)_e)_f$ basis are shown in Fig.~\ref{resources5}.  Referring to Fig.~\ref{U3} for the case $a=1$, we see that in this basis $U_3$ performs a pseudospin rotation about the $z$-axis in both the $f=1/2$ and $3/2$ sectors.  The action of $U_4$ is most easily seen in the $((\blacktriangle\,\blacktriangle)_d\bullet)_f$ basis where, from Fig.~\ref{U4} for the case $b=1$, we know the matrix representation in the $f=1/2$ sector and the $df = \{0\frac{1}{2}, 1\frac{1}{2}\}$ basis is
\begin{eqnarray}
U^{f=1/2}_{4,d}(t) = \left(\begin{array}{cc} 1 &  \\  & e^{-it} \end{array}\right) = e^{-it/2} e^{-it\axis z \cdot \boldsymbol{\sigma}/2};
\end{eqnarray}
and in the $f=3/2$ sector and the $df = \{1 \frac32, 2 \frac32\}$ basis is
\begin{eqnarray}
U^{f=3/2}_{4,d}(t) =  \left(\begin{array}{cc} e^{-it} &  \\  & e^{-it} \end{array}\right) = e^{-it} \mathds{1}.
\end{eqnarray}

To determine the action of $U_4$ on the $f=1/2$ sector in the $(\blacktriangle(\blacktriangle\,\bullet)_e)_f$ basis we once again perform a basis change,
\begin{equation}
(\blacktriangle(\blacktriangle\,\bullet)_e)_{1/2} = \sum_{d} F_{3,ed} ((\blacktriangle\,\blacktriangle)_d\bullet)_{1/2},
\end{equation}
where $F_{3,ed} = \langle ((\blacktriangle\,\blacktriangle)_d\bullet)_{1/2} | (\blacktriangle(\blacktriangle\,\bullet)_e)_{1/2}\rangle$.  The corresponding $2\times 2$ matrix is the same as $F_2$ (a fact which can be understood using the symmetries of the Wigner 6$j$ symbol, see, e.g., Ref.~\onlinecite{Landau}), and generates a basis change from the $d=\{0,1\}$ basis to the $e=\{\frac12,\frac32\}$ basis,
\begin{eqnarray}
	F_3 = \left(\begin{array}{cc} -1/\sqrt{3} & \sqrt{2/3} \\ \sqrt{2/3} & 1/\sqrt{3}\end{array}\right) = \hat{{\bf f}}_2 \cdot \boldsymbol{\sigma}.
	\label{F3}
\end{eqnarray}
It follows that 
\begin{eqnarray}
U^{f=1/2}_{4,e}(t) = F_3 U^{f=1/2}_{4,d}(t) F_3 = e^{-it/2} e^{it \axis n_2 \cdot \boldsymbol{\sigma}/2},
\end{eqnarray}
where $\axis n_2$ is the same rotation axis found in Sec.~\ref{4spins}.  Since in the $f=3/2$ sector $U_4$ is proportional to the identity it will be left unchanged by the basis change to the $(\blacktriangle(\blacktriangle\,\bullet)_e)_{3/2}$ basis,
\begin{eqnarray}
U^{f=3/2}_{4,e}(t) = U^{f=3/2}_{4,d}(t) = e^{-it} \mathds{1}.
\label{U4_3/2}
\end{eqnarray} 

\begin{figure*}
	\begin{center}
		\includegraphics[width=\textwidth]{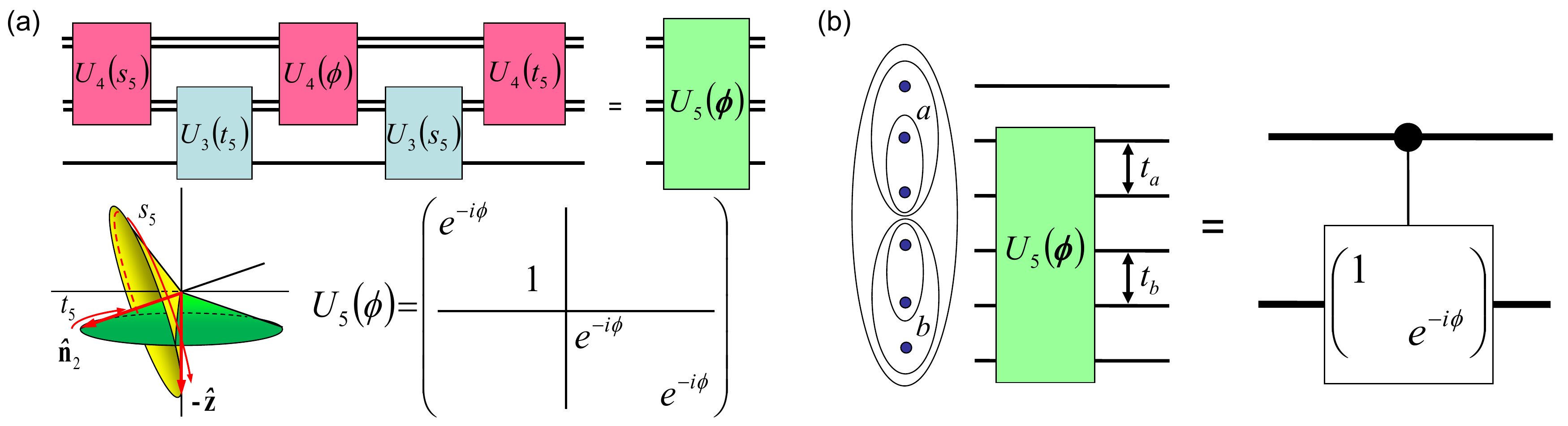}
		\caption{(color online) (a)	Sequence of operations $U_3$ and $U_4$ acting on five spins resulting in the operation $U_5(\phi)$. The sequence is constructed so that the matrix representation of $U_5(\phi)$ is diagonal in the $(\blacktriangle(\blacktriangle\,\bullet)_e)_f$ basis, shown for $ef = \{\frac12\frac12,\frac32\frac12 |\frac12\frac32,\frac32\frac32\}$.  The two-step similarity transformation which carries out this diagonalization is illustrated by the two intersecting cones, where $t_5 = \cos^{-1}(1/4)$ and $s_5 = 2\pi - t_5$.  (b) CPhase gate consisting of $U_5(\phi)$ acting on five spins of two encoded qubits together with two exchange pulses of times $t_a$ and $t_b$ that carry out two single-qubit rotations which depend on the particular choice of short or long $U_3$ sequences.} \label{U5}
	\end{center}
\end{figure*}

At this point we are ready to complete our two-qubit gate construction.  To do this, we need to produce a sequence of operations acting on five out of the six spins forming the two encoded qubits in states $a$ and $b$ (see Fig.~\ref{twoqubits}(a)) which applies a phase factor of $e^{-i\phi}$ to the state with $ab=11$.  To see what is required note that the two-qubit state $|1_L\rangle|1_L\rangle$ can be expressed as $((\bullet\,\blacktriangle)_{1/2}(\blacktriangle\,\bullet)_{1/2})_g$ where $g$ equals 0 or 1.  It is straightforward to expand these states as follows,
\begin{eqnarray}
((\bullet\,\blacktriangle)_{\frac12}(\blacktriangle\,\bullet)_{\frac12})_0 &=& (\bullet(\blacktriangle(\blacktriangle\,\bullet)_{\frac12})_{f=\frac12})_0,\label{f12}\\
((\bullet\,\blacktriangle)_{\frac12}(\blacktriangle\,\bullet)_{\frac12})_1 &=& 
-\frac13 (\bullet(\blacktriangle(\blacktriangle\,\bullet)_{\frac12})_{f=\frac12})_1\nonumber \\
&&+
\frac{2\sqrt{2}}{3}(\bullet(\blacktriangle(\blacktriangle\,\bullet)_{\frac12})_{f=\frac32})_1,
\label{f32}
\end{eqnarray}
where in \eqref{f32} we have used the recoupling coefficients $F_{4,\frac12 f} = \langle (\bullet\,(\blacktriangle\,\bullet)_f)_1  | ((\bullet\,\blacktriangle)_{1/2}\,\bullet)_1 \rangle$ where $F_{4,\frac12 \frac12} = -1/3$, $F_{4,\frac12 \frac32} = 2\sqrt{2}/3$.  Here the rightmost $\bullet$ in the definition of $F_4$ represents the rightmost qubit in the state $(\blacktriangle\,\bullet)_{1/2}$ in \eqref{f32}.  To apply a phase factor of $e^{-i\phi}$ to both states on the left-hand sides of \eqref{f12} and \eqref{f32} it is clearly necessary to apply this same phase factor to the five-spin states $(\blacktriangle(\blacktriangle\,\bullet)_{1/2})_{1/2}$ and $(\blacktriangle(\blacktriangle\,\bullet)_{1/2})_{3/2}$.  We therefore need to find a sequence of operations that produces an operation diagonal in the $(\blacktriangle(\blacktriangle\,\bullet)_e)_f$ basis.

The simplest such diagonal operation is produced by a single action of the operation $U_3$.  However, as can be seen in Fig.~\ref{resources5}, this operation applies a different phase factor to the states $(\blacktriangle(\blacktriangle\,\bullet)_{1/2})_{f=1/2}$ and $(\blacktriangle(\blacktriangle\,\bullet)_{1/2})_{f=3/2}$.   It is then natural to again try to apply the three-operation construction $U_4(t)U_3(\bar t)U_4(t)$ of Sec.~\ref{3spins}.  However, as in Sec.~\ref{4spins}, this construction is incapable of producing the required operation.  Direct calculation shows that it is impossible to produce an operation for which the same nontrivial phase factor is applied to the states $(\blacktriangle(\blacktriangle\,\bullet)_{1/2})_f$ with $f=1/2$ and $f=3/2$.  Performing four operations, i.e. a sequence of the form $U_3 U_4 U_3 U_4$, is equivalent to $U_4 U_3 U_4$ because the final $U_3$ operation is a single-qubit rotation.  We must therefore consider a sequence of at least five operations, and the explicit construction presented below shows that five is indeed enough.

The sequence shown in Fig.~\ref{U5}(a) is designed to multiply the two states $(\blacktriangle(\blacktriangle\,\bullet)_{1/2})_{f=1/2,\, 3/2}$ by the same phase factor of $e^{-i\phi}$.  The sequence has the form $U_5(\phi) = U_4(s_5) U_3(t_5) U_4(\phi) U_3(s_5) U_4(t_5)$ where $s_5 = 2\pi - t_5$ so that $U_4(s_5) = U_4(t_5)^{-1}$ and $U_3(s_5) = U_3(t_5)^{-1}$ in the $ab=11$ Hilbert space.  Similar to $U_4$ in Sec.~\ref{4spins}, in this space the $U_5$ construction has the structure of a similarity transformation, $U_5(\phi) = S U_4(\phi) S^{-1}$ with $S = U_4(s_5) U_3(t_5)$.  In both the $f=1/2$ and $f=3/2$ sectors this similarity transformation can be visualized as a series of pseudospin rotations.

Again referring to Fig.~\ref{resources5} for the case of $f=3/2$, $U_4(\phi)$ is equal to the identity times $e^{-i\phi}$.  This immediately implies that in this sector the similarity transformation has no effect.  Thus, in the $f=3/2$ sector, $U_5(\phi)$ equals $U_4(\phi)$ and, in particular, multiplies the state $(\blacktriangle(\blacktriangle\,\bullet)_{1/2})_{3/2}$ by $e^{-i\phi}$. 

To understand the action of $U_5(\phi)$ on the $f=1/2$ sector, note that in this sector $U_4(\phi)$ is a pseudospin rotation about the axis $\axis n_2$.  In order for $U_5(\phi)$ to multiply the state $(\blacktriangle(\blacktriangle\,\bullet)_{1/2})_{1/2}$ by $e^{-i\phi}$, the similarity transformation carried out by $S$ in this sector must be chosen so that it rotates $\axis n_2$, the rotation axis of $U_4(\phi)$,  to $-\axis z$.  As shown in Fig.~\ref{U5}, $S$ consists of a rotation about the $z$-axis through the angle $t_5$ (green cone) followed by a rotation about the $n_2$-axis through the angle $s_5 = 2\pi -t_5$ (yellow cone).  It is straightforward to show that if we choose
\begin{equation}
	t_5 = \cos^{-1} \frac{\axis n_2\cdot \axis z}{\axis n_2\cdot \axis z - 1} = \cos^{-1}\frac14,
\end{equation}
then, under these rotations, $\axis n_2$ is first rotated to the intersection of the green and yellow cones, and then rotated to $-\axis z$.

\begin{figure*}%
	\centering
	\includegraphics[width=\textwidth]{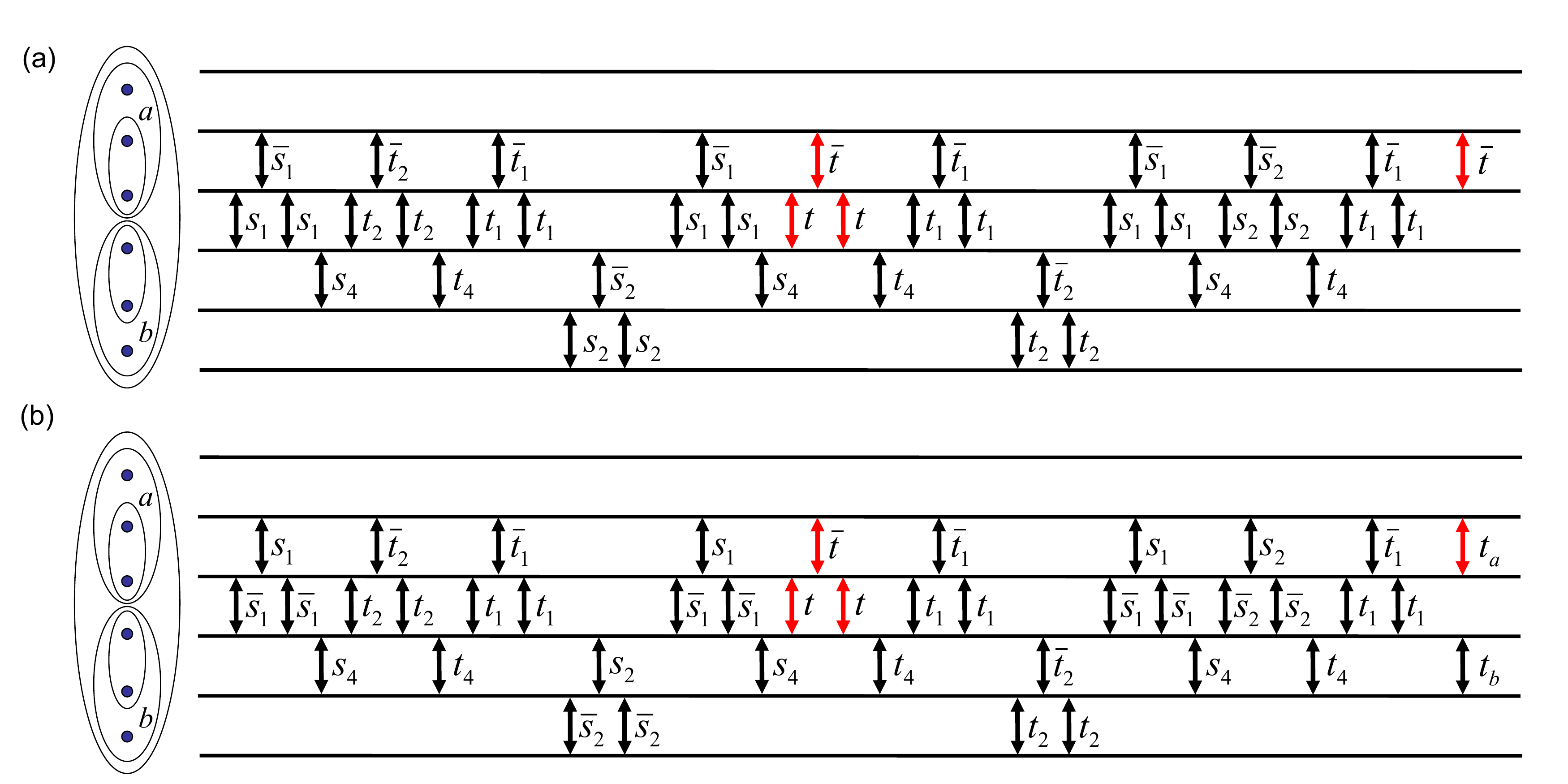}
	\caption{(color online) Two full pulse sequences of 39 pulses plus single qubit rotations for the CPhase gate construction shown in Fig.~\ref{U5}(b): (a) a 40-pulse sequence (with one single-qubit rotation); and (b) a 41-pulse sequence (with two single-qubit rotations) of slightly shorter total duration obtained by swapping pulse labels $s_1 \leftrightarrow \bar s_1$ and $s_2 \leftrightarrow \bar s_2$.  Red pulses depend on the phase $\phi$ which determines the CPhase gate, and  black pulses are independent of $\phi$. Ignoring single-qubit rotations, the pulse times which do not depend on $\phi$ are $t_4 = 2\pi/3$, $s_4 = 4\pi/3$, as well as $t_1=1.34004$ and $t_2 = 0.86463$ (obtained by solving for the short sequences for $U_3(t_4)$ and $U_3(t_5 = \cos^{-1}(1/4))$, respectively, see Sec.~\ref{3spins}), together with those obtained from the relations $\tan(t_i/2)\tan(\bar t_i/2)=-2$ and $\bar t_i + \bar s_i = t_i+s_i=2\pi$ for $i=1,2$. The times $t$ and $\bar t$ depend on $\phi$ and are found by solving for the short or long sequence for $U_3(\phi)$.  In (a) a single pulse of duration $\bar t$ brings the final gate to CPhase form, whereas in (b) two pulses acting on both qubits are required, with pulse durations $t_a= 4.11499 + \bar t\ ({\rm mod}\ 2\pi)$ and $t_b=2.73045$.    For $\phi = \pi$ the CPhase gate carried out by these pulse sequences is locally equivalent to CNOT and, if we choose the short sequence for the central $U_3(\phi = \pi)$, we find $t = 1.91063$ and $\bar t = 4.37255$.}%
	\label{39steps}%
\end{figure*}

The outcome of this transformation in the $f=1/2$ sector of $U_5(\phi)$ in the $(\blacktriangle(\blacktriangle\,\bullet)_e)_{1/2}$ basis with $e=\{\frac12,\frac32\}$ is
\begin{eqnarray}
U_{5,e}^{f=1/2}(\phi) = e^{-i\phi/2} e^{i\phi (-\axis z) \cdot\boldsymbol{\sigma}/2} = \left(\begin{array}{cc} e^{-i\phi} & \\ & 1 \end{array}\right).
\end{eqnarray}
Thus the state with $((\bullet\,\bullet)_{a=1}((\bullet\,\bullet)_{b=1}\,\bullet)_{1/2})_{1/2}$ is multiplied by a phase factor of $e^{-i\phi}$. As shown above, in the $f=3/2$ sector $U_5(\phi)$ is proportional to the identity and multiplies the state $((\bullet\,\bullet)_{a=1}((\bullet\,\bullet)_{b=1}\,\bullet)_{1/2})_{3/2}$ by the same phase factor of $e^{-i\phi}$.  So the action of $U_5(\phi)$ is to multiply all states with $g=0$ and 1 on the right-hand sides of \eqref{f12} and \eqref{f32} by $e^{-i\phi}$.  

The resulting operation is thus locally equivalent to a CPhase gate. To complete the gate construction we need only determine the single-qubit rotations needed to set the $a=0$ phase factor, discussed in Sec.~\ref{4spins}, and the $b=0$ phase factor, discussed above, to 1.  The value of these phase factors depend on whether we use short sequences or long sequences for the $U_3$ operations throughout the construction.  Whatever the value of these phase factors, they can be set to 1 by performing single-qubit rotations corresponding to the two pulses shown in Fig.~\ref{U5}(b).   

Before proceeding we point out that any sequence of exchange pulses that acts on only five spins (which we take to be the five rightmost in Fig.~\ref{twoqubits} with total spin $f$) and that carries out a leakage-free two-qubit gate in the total spin $g=1$ sector, must carry out the same gate in the total spin $g=0$ sector. This is because i) any such sequence conserves $f$; and ii) for both $g=0$ and $g=1$ the two-qubit basis states with $ab=00,01,10,$ and 11 all have nonzero projection onto the $f=1/2$ sector.  [For $g=0$, $f$ is fixed to be 1/2.  For $g=1$, when $a=0$, $f$ is also fixed to be 1/2, and, when $a=1$, the expansion \eqref{f32}, together with a similar expansion for the case $b=0$ with the same recoupling coefficients, implies nonzero projection onto the $f=1/2$ sector.]  Any operation produced by a pulse sequence which acts on the five rightmost spins will then have identical matrix representations in two two-qubit subspaces with the same $ab$ basis choice: one in the $g=0$ sector, which lives entirely in the $f=1/2$ sector, and another in the $g=1$ sector, after projection onto the $f=1/2$ sector.  Therefore if this sequence produces a leakage-free two-qubit gate in the $g=1$ sector it will produce the same leakage-free two-qubit gate in $g=0$ sector.  This observation is consistent with the fact that the Fong-Wandzura sequence\cite{fong11} and related sequences in Ref.~\onlinecite{setiawan13}, as well as our sequences, act on only five spins, while the sequence of Ref.~\onlinecite{divincenzo00} acts on six.

\section{Full Pulse Sequences}
\label{full}

Figure \ref{39steps} shows two explicit pulse sequences for CPhase gates obtained by unpacking the $U_4$ and $U_3$ operations in Fig.~\ref{U5} and replacing them with sequences of exchange pulses.  To do this unpacking, each $U_3$ operation, including those within each $U_4$ operation, are replaced by three-pulse sequences found by solving \eqref{ttbar} and \eqref{phit} (see Sec.~\ref{3spins}).  To determine the pulse times for the entire sequence it is necessary to solve these equations for $U_3(x)$ when $x = t_4, t_5, s_4, s_5$  (see Secs.~\ref{4spins} and \ref{5spins}), as well as $x=\phi$ where $\phi$ is the phase which characterizes the CPhase gate.  The two sequences shown in Fig.~\ref{39steps} correspond to different choices for the two possible three-pulse sequences that can be used to carry out each $U_3$ operation, the short sequence and the long sequence. 

In Fig.~\ref{39steps}(a) each $U_3(t_4)$ and $U_3(t_5)$ are taken to be short sequences, while each $U_3(s_4)$ and $U_3(s_5)$ are taken to be long sequences.  For this choice $U_3(s_4) = U_3(t_4)^{-1}$ and $U_3(s_5) = U_3(t_5)^{-1}$ in the full Hilbert space, not just in the $ab=11$ subspace.  As a consequence, from the palindromic form of the full sequence, it is apparent that the $a=0$ phase factors contributed by those $U_3$ operations which act on the two spins in the state $a$ cancel, save that due to the single $U_3(\phi)$ in the center of the sequence for $U_4(\phi)$, which is itself at the center of the sequence for $U_5(\phi)$.  This phase factor is eliminated by the single-qubit rotation carried out by the single red pulse at the end of the sequence.  The $b=0$ phase factors contributed by $U_3(s_5)$ and $U_3(t_5)$ in Fig.~\ref{U5}(a) cancel completely and there is no need for a single-qubit rotation on the qubit in the state $b$.  All of the pulse times are fixed except for the four pulses shown in red: three in the center, with times labeled $t$, $\bar t$, and $t$, which carry out $U_3(\phi)$, and the one at the end of the sequence mentioned above of time $\bar t$ which removes the $a=0$ phase factor.  The pulse times, including those for the $\phi$-dependent red pulses when $\phi=\pi$, are given explicitly in the figure caption.  

\begin{figure*}%
	\includegraphics[width=\textwidth]{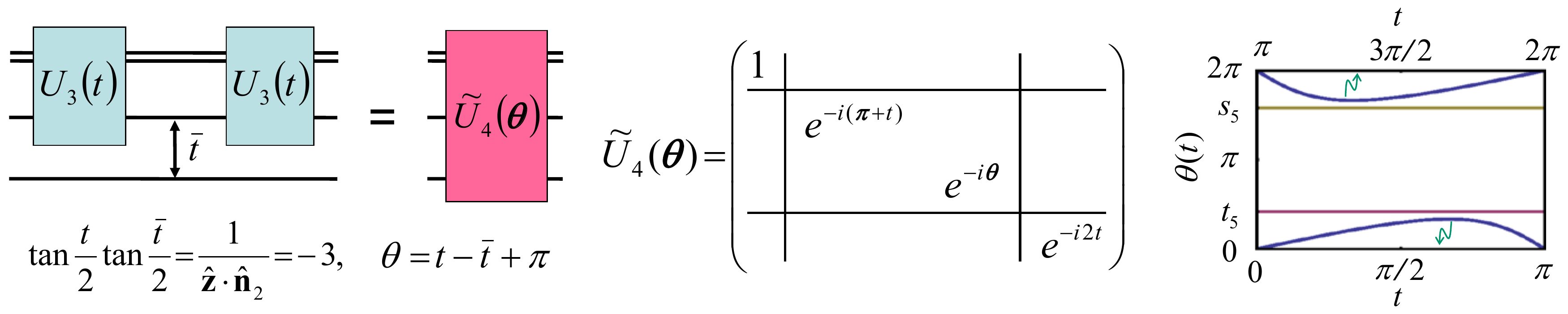}%
	\caption{(color online) Sequence of two $U_3(t)$ operations and one exchange pulse, $U_2(\bar t)$, which produces the four-spin operation $\widetilde U_4(\theta)$.  The sequence is based on the same geometric principle as that for $U_3$ (see Sec.~\ref{4spins} and Fig.~\ref{U3}).  The matrix representation of $\widetilde U_4(\theta)$ is diagonal in the $(\blacktriangle (\bullet\,\bullet)_b)_d$ basis, shown for $bd = \{10|01,11|12\}$, and induces a phase difference $\theta$ between the states with $bd = 10$ and 11.  To use $\widetilde U_4$ in the construction of $U_5$ (see Sec.~\ref{5spins}) it would be necessary to choose $t$ and $\bar t$ so that $\theta = t_5 = \cos^{-1} 1/4$ and $s_5 = 2\pi - t_5$.  However, the graph of $\theta$ vs.\ $t$ shows that this cannot be achieved with a single $\widetilde U_4$ operation. }
	\label{U4tilde}%
\end{figure*}

In Fig.~\ref{39steps}(b) we continue to take each $U_3(t_4)$ and $U_3(t_5)$ to be short sequences, but now also take each $U_3(s_4)$ and $U_3(s_5)$ to be short sequences. The resulting full sequence then has slightly shorter total duration than that shown in Fig.~\ref{39steps}(a) provided all pulses are performed in parallel when possible. Switching to only short sequences amounts to swapping the pulses of length $s_i$ with those of length $\bar s_i$ for $i=1$ and 2 in Fig.~\ref{39steps}(a).  The price of this rearrangement is that we lose the phase factor cancellations which occurred in the first sequence.   Because of this, two single-qubit rotations corresponding to the two pulses at the end of the sequence, rather than just one, are needed to eliminate the $a=0$ and $b=0$ phase factors. One pulse acts on the two spins in the state $a$ for time $t_a$ which depends on $\phi$ through $\bar t$, and the second pulse acts on the two spins in the state $b$ for time $t_b$ which is independent of $\phi$.  Both $t_a$ (as a function of $\bar t$) and $t_b$ are given in the figure caption.

Our 39-pulse sequence is significantly longer than both the 19-pulse DiVincenzo et al.\cite{divincenzo00} sequence (which only carries out a gate locally equivalent to CNOT in the total spin $g=1$ sector) and the 18-pulse Fong and Wandzura\cite{fong11} sequence (which, like our sequence, carries out a gate locally equivalent to CNOT in both the total $g=0$ and $g=1$ sectors) as well as the related 16- and 14-pulse sequences for geometries other than linear arrays.\cite{setiawan13}  Nevertheless, we believe our two-qubit gate construction is of interest, because it introduces new methods for finding pulse sequences acting on large Hilbert spaces by effectively reducing the size of this Hilbert space at each stage of the construction.

\section{Conclusions}

We have presented an analytic construction of pulse sequences for exchange-only quantum computation which carry out entangling, leakage-free two-qubit gates on qubits encoded using three spin-1/2 particles.  The resulting pulse sequences, while far from the most efficient, have the unique property that they can be understood in simple geometric terms despite the enormous size of the space of unitary operators acting on the full Hilbert space of the six spin-1/2 particles needed to encode two qubits.  The essential idea behind our construction is that this Hilbert space can be built up, spin-by-spin, in such a way that at each level---from two spins, to three, then four, and finally five spins---we are able to reduce the relevant effective Hilbert spaces to either trivial one-dimensional sectors or two-dimensional sectors which can be visualized in the language of spin-1/2 pseudospins. 

Because each level of our construction can be understood in terms of effective spin-1/2 pseudospins we are able to work out the required pulse sequences analytically, without having to resort to numerical minimization of a cost function (as in Ref.~\onlinecite{divincenzo00}), the use of genetic algorithms (as in Ref.~\onlinecite{fong11}), or any other numerical method.  In addition, because our construction is analytic it allows us to envision alternate pulse sequences for carrying out two-qubit gates, some of which are discussed in Appendix \ref{alternate}.  We believe this general approach of iteratively constructing pulse sequences acting on large Hilbert spaces by effectively reducing the size of the Hilbert space at each level of iteration may have wider applicability for constructing useful pulse sequences for quantum computation.

\acknowledgments

DZ and NEB thank Guido Burkard for useful discussions.  This work is supported by US DOE Grant No. DE-FG02-97ER45639.

\appendix

\section{Alternate $U_4$ Construction}
\label{alternate}

In Sec.~\ref{4spins} we introduced the operation $U_4$, which was then used as a building block of $U_5$ in our full CPhase gate construction.  As shown in Fig.~\ref{resources4}, the operation $U_4$ was, itself, constructed out of a sequence of three $U_3$ operations and two $U_2$ operations. 

As emphasized in Sec.~\ref{4spins}, an important feature of $U_4$ is that it is diagonal in the $((\bullet\,\bullet)_{a=1}(\bullet\,\bullet)_b)_d \rightarrow (\blacktriangle(\bullet\,\bullet)_b)_d$ basis (as in the main text, $\blacktriangle$ is an effective spin-1 particle which here corresponds to the two spin-1/2 particles with total spin $a=1$ shown in Fig.~\ref{resources4}).  It is natural to ask if we can use the three-pulse sequence of Sec.~\ref{3spins} to construct an alternate $U_4$ operation which is also diagonal in this basis, but which only requires two $U_3$ operations and a single exchange pulse, $U_2$.  The answer is yes, but, as shown below, the resulting operation cannot be directly used in our CPhase gate construction.

Figure \ref{U4tilde} shows this alternate $U_4$ construction.  We denote the resulting operation $\widetilde U_4$.  As in Sec.~\ref{4spins}, the only two-dimensional sector in the $(\blacktriangle(\bullet\,\bullet)_b)_d$ basis is that with $d=1$ and we again define a pseudospin with $\uparrow = (\blacktriangle(\bullet\,\bullet)_0)_1$ and $\downarrow = (\blacktriangle(\bullet\,\bullet)_1)_1$.  As shown in Fig.~\ref{resources4}, the operations $U_3$ and $U_2$ are then pseudospin rotations about the $\axis n_2$- and $\axis z$- axes, respectively.  

The specific three-operation sequence $U_3(t)U_2(\bar t)U_3(t)$ used to construct $\widetilde U_4$ is found using the same geometric construction used for $U_3$ in Sec.~\ref{3spins}.  The only difference is that the two rotation axes make a different angle than in the $U_3$ construction. This alters the right-hand side of \eqref{ttbar} with $1/{\axis n_1}\cdot{\axis z} = -2$ replaced by $1/{\axis n_2}\cdot {\axis z} = -3$, as shown in Fig.~\ref{U4tilde}. Nevertheless, provided this modified form of \eqref{ttbar} is satisfied, the resulting operation will still be a $z$-axis pseudospin rotation, and hence diagonal in the $(\blacktriangle(\bullet\,\bullet)_b)_d$ basis.

Direct calculation gives the full matrix representation for $\widetilde U_4$ shown in Fig.~\ref{U4tilde}.  For our construction, the crucial phase difference is that between the $bd = 01$ and the $bd=11$ diagonal matrix elements (see below), which we denote $\theta$, and which is related to $t$ and $\bar t$ through the relation $\theta = t - \bar t + \pi$, also given in the figure caption.  

To see why the phase difference $\theta$ is important, consider the construction of $U_5$ in Sec.~\ref{5spins}.  The sequence of operations used in this construction is $U_5(\phi) = U_4(s_5) U_3(t_5) U_4(\phi) U_3(s_5) U_4(t_5)$.  In this sequence the two outermost $U_4$ operations ($U_4(t_5)$ and $U_4(s_5)$) perform rotations about the axis $\axis n_2$ through the angles $t_5 = \cos^{-1}1/4$ and $s_5 = 2\pi - t_5$, respectively, on the pseudospin space with $\uparrow_{f} = (\blacktriangle(\blacktriangle\,\bullet)_{1/2})_{f}$ and $\downarrow_{f} = (\blacktriangle(\blacktriangle\,\bullet)_{3/2})_{f}$ for $f=1/2$ (referring to Fig.~\ref{resources5}).  As shown in Sec.~\ref{5spins}, the pseudospin rotation angle produced by $U_4$ in the $f=1/2$ sector is equal to the phase difference between the $bd = 10$ and $bd =11$ diagonal matrix elements which, in Fig.~\ref{U4}, is denoted $\phi$.  This phase can be set to any desired value since it is determined by the phase $\phi$ which appears in the central $U_3(\phi)$ operation.  As noted in Sec.~\ref{5spins}, in the $f=3/2$ pseudospin space the central $U_4(\phi)$ operation is proportional to the identity and so is unchanged by the similarity transformation construction which produces $U_5(\phi)$.

Unfortunately, we cannot directly replace each of the two outer $U_4$ operations in the $U_5$ construction with $\widetilde U_4$ operations.  To do so, it would be necessary to choose $t$ and $\bar t$ so that $\theta = t_5$ (and $\theta = s_5 = 2\pi - t_5$); however, as shown in Fig.~\ref{U4tilde}, in contrast to $\phi$ for $U_4$, the range of achievable $\theta$ values does not include $t_5$ (or $s_5$).  Note that we do not consider replacing the central $U_4$ with $\widetilde U_4$ because, in our construction, it is crucial that this operation be proportional to the identity in the $f=3/2$ sector, and this is not the case for $\widetilde U_4$.

However, we {\it can} replace each outer $U_4$ operation with products of two $\widetilde U_4$ operations. This is because, as can be seen in Fig.~\ref{U4tilde}, the continuum of achievable $\theta$ values includes $t_5/2$ (and $s_5/2)$.  Thus there are a continuum of products of the form $\widetilde U_4(\theta_2) \widetilde U_4(\theta_1)$ where $\theta_1 + \theta_2 = t_5$ (and $\theta_1 + \theta_2 = s_5$) which will perform the required pseudospin rotations in the $f=1/2$ sector for our $U_5$ construction.  Since each $\widetilde U_4$ operation is realized through a sequence of the form $U_3 U_2 U_3$, the product of two $\widetilde U_4$ operations will always have the form $U_3 U_2 U_3 U_2 U_3$, where the two central $U_3$ operations are combined into a single $U_3$.  These $\widetilde U_4(\theta_2) \widetilde U_4(\theta_1)$ product operations thus contain precisely the same number of pulses as the $U_4$ operations constructed in Sec.~\ref{4spins} and can be used to construct a continuum of full 39-pulses sequences (plus two pulses for single-qubit rotation) which carry out CPhase gates.

\begin{figure}%
	\includegraphics[width=\columnwidth]{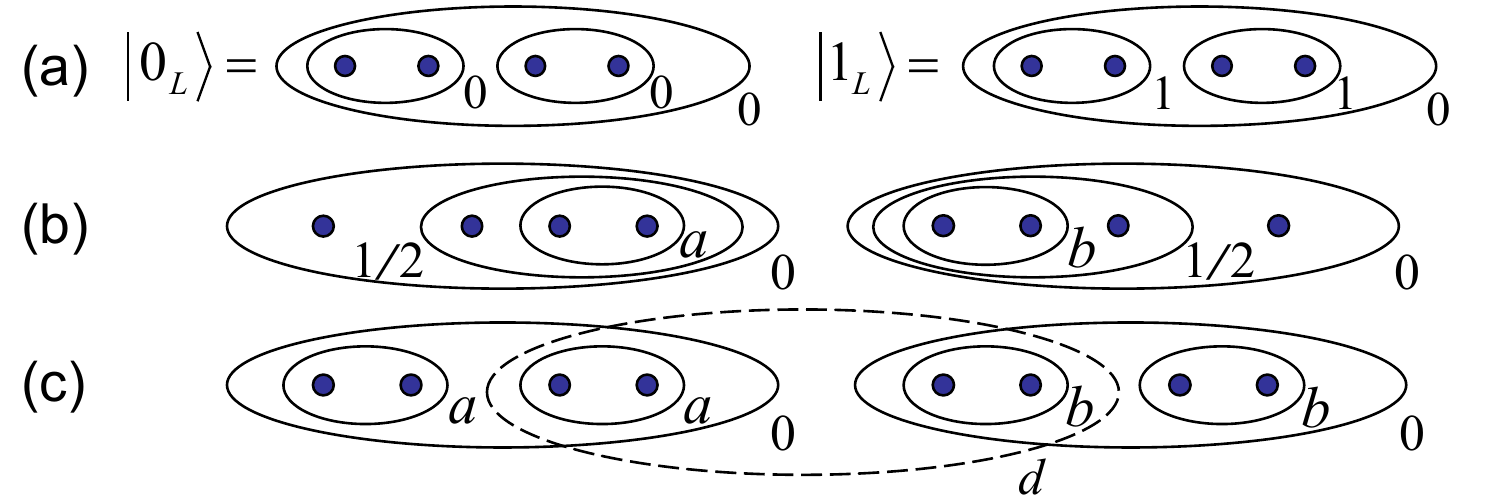}%
	\caption{(a) Qubit encoding using four spin-1/2 particles.   (b) Two four-spin qubits in states $a$ and $b$, expressed in a basis which shows that if the outer two spins are ignored the remaining spins form two three-spin qubits as in Fig.~\ref{twoqubits}(a).  (c) The same four-spin qubits with a dashed oval enclosing the four central spins labeled by total spin $d$.  In the text we show that no sequence of exchange pulses acting only on these four central spins can result in an entangling two-qubit gate.}%
	\label{proof}%
\end{figure}

\section{Four Spins are Not Enough}
\label{five}

Here we show that any sequence of exchange pulses which carries out a leakage-free, entangling two-qubit gate on two three-spin qubits, independent of whether the total spin of all six particles is 0 or 1, must act on at least five spins.\cite{proof_note}   

To do this it is convenient to consider logical qubits encoded using four spins rather than just three.  This four-spin encoding is shown in Fig.~\ref{proof}(a) (the noncomputational states are those for which the total spin of the four spin-1/2 particles is 1 or 2).  Figure \ref{proof}(b) shows two adjacent four-spin qubits, and illustrates the fact that if we remove the two outermost spins the remaining three spins in each logical qubit have total spin 1/2 and are therefore precisely the three-spin qubits used in our main construction.  Thus any pulse sequence which performs a two-qubit gate on two three-spin qubits regardless of whether their total spin is 0 or 1 must carry out the same two-qubit gate on two four-spin qubits when acting on the six central spins in Fig.~\ref{proof}(b).

It follows that the Fong-Wandzura sequence, as well as our sequences, can be used to carry out two-qubit gates on pairs of four-spin qubits.  We note that a 34-pulse sequence which produces a two-qubit gate locally equivalent to CNOT for two four-spin qubits was also found numerically in Ref.~\onlinecite{hsieh03} using methods similar to those used in Ref.~\onlinecite{divincenzo00}. However, this sequence acts on all eight spins used to encode the two qubits and thus cannot be used to carry out two-qubit gates for three-spin qubits. 

Now consider an operation produced by exchange pulses which only act on the four central spins, i.e. those circled by the dashed line labeled by the total enclosed spin $d$, in Fig.~\ref{proof}(c).  We denote the resulting unitary operation $U^{(4)}$.  If we assume that $U^{(4)}$ carries out a leakage-free two-qubit gate then it is clear that $U^{(4)}$ must be diagonal in the $((\bullet\,\bullet)_a (\bullet\,\bullet)_b)_d$ basis.  If this were not the case then either $a$, $b$, or both would change after carrying out $U^{(4)}$.  As a result, one or both of the four-spin qubits would undergo a transition to a state in which the two pairs of spins within the qubit have different total spin values.  Since any such four-spin state cannot have total spin 0, this transition would lead to leakage out of the encoded four-spin qubit space shown in Fig.~\ref{proof}(a).

Since $U^{(4)}$ is diagonal in the $((\bullet\,\bullet)_a (\bullet\,\bullet)_b)_d$ basis it must give each encoded two-qubit state in the $ab$ basis a phase factor, $e^{i\phi_{ab}}$.  For the two-qubit states with $ab=00, 01,$ and 10 the value of the total spin $d$ is fixed to be 0, 1, and 1, respectively, and so the corresponding phase factors are single elements in the matrix representation of $U^{(4)}$.  However, for the case $ab=11$ the value of $d$ can be either 0, 1 or 2.  Moreover, states with all three $d$ values have non-zero overlap with two four-spin qubits in the $ab=11$ state.  To see this, first express this state, $\big(((\bullet\,\bullet)_{a=1}(\bullet\,\bullet)_{a=1})_0((\bullet\,\bullet)_{b=1}(\bullet\,\bullet)_{b=1})_0\big)_0$, as $((\blacktriangle\,\blacktriangle)_0(\blacktriangle\,\blacktriangle)_0)_0$ (where, as in the main text, $\blacktriangle$ is an effective spin-1 particle).  This state can then be expanded in basis states with well-defined $d$ quantum numbers as follows,
\begin{eqnarray}
&&((\blacktriangle\,\blacktriangle)_0(\blacktriangle\,\blacktriangle)_0)_0 = (\blacktriangle\,( \blacktriangle (\blacktriangle\,\blacktriangle)_0)_1)_0 \nonumber \\
&&~~~~=\frac13 (\blacktriangle\,((\blacktriangle\,\blacktriangle)_{d=0}\blacktriangle)_1)_0 
- \frac{1}{\sqrt{3}} (\blacktriangle\,((\blacktriangle\,\blacktriangle)_{d=1}\blacktriangle)_1)_0\nonumber\\
&&~~~~~~~~~~~~~~~~~~~+\frac{\sqrt{5}}{3} (\blacktriangle\,((\blacktriangle\,\blacktriangle)_{d=2}\blacktriangle)_1)_0,
\end{eqnarray}
where we have used the recoupling coefficients $F_{5,0d} = \langle ((\blacktriangle\,\blacktriangle)_d\,\blacktriangle)_1 | (\blacktriangle\,(\blacktriangle\,\blacktriangle)_0)_1 \rangle$ where $F_{5,00} = 1/3$, $F_{5,01} = -1/{\sqrt 3}$, and $F_{5,02} = \sqrt5/3$.  Since these coefficients are non-zero for all possible values of $d$, the phase factor, $e^{i\phi_{11}}$, produced by $U^{(4)}$ for the state $ab=11$ must be the same for $d = 0$, 1, and 2.

The above discussion shows that in the $((\bullet\,\bullet)_a(\bullet\,\bullet)_b)_d$ basis with $abd=\{000,110|011,101,111|112\}$, the matrix representation of $U^{(4)}$ must have the form
\begin{equation}
U^{(4)}
	=
	\left(
	\begin{array}{cc|ccc|c}
	e^{i\phi_{00}}&&&&&\\
	&e^{i\phi_{11}}&&&&	\\\hline
	&&e^{i\phi_{01}}&&&\\
	&&&e^{i\phi_{10}}	&\\
	&&&&e^{i\phi_{11}}	\\\hline
	&&&&&e^{i\phi_{11}}
	\end{array}
	\right).
	\label{U4matrix}
\end{equation}
It is straightforward to show that the two-qubit gate produced by $U^{(4)}$ is locally equivalent to a controlled rotation through the angle $\phi = \phi_{00} - \phi_{01} -\phi_{10} + \phi_{11}$.  The requirement that $U^{(4)}$ produce an entangling two-qubit gate is then
\begin{equation}
	\phi_{00} - \phi_{01} - \phi_{10} + \phi_{11} \neq 0 \qquad (\text{mod}~2\pi).
	\label{basic_phases}
\end{equation}

We denote the determinant of $U^{(4)}$ in a sub-sector of total spin $d$ as $\det U^{(4)}\vert_{d}$.  Equations \eqref{U4matrix} and \eqref{basic_phases} then imply the following condition on $U^{(4)}$,
\begin{equation}
	\frac{\det U^{(4)}\vert_{d=0} \det U^{(4)}\vert_{d=2}}{\det U^{(4)}\vert_{d=1}} \neq 1.
	\label{determinant1}
\end{equation}
If $U^{(4)}$ is the result of a series of $N$ exchange pulses it has the form
\begin{equation}
	U^{(4)} = U_N(t_N) \cdots U_2(t_2) U_1(t_1).
	\label{product}
\end{equation}
Here $U_n(t_n)=\exp(-i H_n t_n)$ is the time evolution operator of the $n$-th pulse, where $H_n =\mathbf S_{i(n)}\cdot \mathbf S_{j(n)} + \frac34$ is the Hamiltonian of the Heisenberg exchange interaction between spins $i$ and $j$ with the constant added for convenience.  

Since the determinant of a product of operators is equal to the product of the determinants of those operators, the requirement that the condition \eqref{determinant1} hold for the sequence \eqref{product} implies that
\begin{equation}
 \frac{\det U_n\vert_{d=0} \det U_n\vert_{d=2}}{\det U_n\vert_{d=1}} \neq 1,
	\label{determinant2}
\end{equation}
for at least one of the $U_n$ operations.  Given that $\det U_n=\det e^{-it_n H_n}=e^{-it_n \text{Tr}\ H_n}$, this condition can be translated into a condition on the trace of the Hamiltonian of a single pulse.  If we denote the trace of $H_n$ within a sector of total spin $d$ as $\text{Tr}\ H_n \vert_{d}$, then \eqref{determinant2} implies that at least one Hamiltonian $H_n$ pulsed in \eqref{product} must satisfy the condition
\begin{equation}
	\text{Tr}\ H_n \vert_{{d=0}} - \text{Tr}\ H_n \vert_{{d=1}} + \text{Tr}\ H_n \vert_{{d=2}} \neq 0.
	\label{eq:trace}
\end{equation}
However, for $H_n = {\bf S}_{i(n)} \cdot {\bf S}_{j(n)}+\frac34$ where spins $i(n)$ and $j(n)$ label two of the four central spins in Fig.~\ref{proof}(c), one finds that $\text{Tr}\ H_n \vert_{d=0} = 1$, $\text{Tr}\ H_n \vert_{d=1} = 2$, and $\text{Tr}\ H \vert_{d=2} = 1$.  Thus we see that
\begin{equation}
	\text{Tr}\ H_n \vert_{{d=0}} - \text{Tr}\ H_n \vert_{{d=1}} + \text{Tr}\ H_n \vert_{{d=2}} = 1 -2 + 1 = 0.
	\label{trace2}
\end{equation}
It immediately follows that any pulse sequence consisting of exchange pulses between two of the four central spins in Fig.~\ref{proof}(c) cannot produce an operation of the form \eqref{U4matrix} and thus cannot produce a leakage-free, entangling two-qubit gate.

Lastly, we point out that if the trace condition \eqref{trace2} holds for two operators, $H_1$ and $H_2$, it trivially also holds for their sum $H_1 + H_2$.  It immediately follows that our result that acting on only four spins is not sufficient to carry out an entangling two-qubit gate holds not just when the exchange interaction is pulsed in series, but also when it is pulsed in parallel (e.g. when operations of the form $e^{-it({\bf S}_1 \cdot {\bf S}_2  + {\bf S}_2 \cdot {\bf S}_3)}$ are included).  This also follows from the fact that such parallel operations can always be approximated, to any required accuracy, by sequences of operations carried out in series, as shown in Ref.~\onlinecite{kempe01}.

\bibliography{bibliography}

\begin{thebibliography}{26}
\bibitem [{sig()}]{sign_convention}%
  \BibitemOpen
  \href@noop {} {}\bibinfo {note} {Since we do not use wave functions written
  explicitly in terms of $S_z$ quantum numbers it is necessary to specify a
  phase convention for states labeled only by total spin quantum numbers. We
  take this convention to be the standard Condon-Shortley phase choice. This
  choice also determines the phases of the relevant spin-recoupling
  coefficients. It is because these coefficients are real-valued that all
  pseudospin rotation axes in our construction lie in the
  $xz$-plane.}\BibitemShut {Stop}%
\bibitem [{int()}]{integrated_pulse}%
  \BibitemOpen
  \href@noop {} {}\bibinfo {note} {Actual pulses will not simply involve
  switching on and off a constant exchange coupling $J$, and the dimensionless
  pulse duration $t$ is more accurately taken to be the time integral of $J(t)$
  over the pulse. For our choice of units, $t=\pi$ corresponds to a SWAP
  operation.}\BibitemShut {Stop}%
\bibitem [{ttb()}]{ttbart}%
  \BibitemOpen
  \href@noop {} {}\bibinfo {note} {It is interesting to note that three pulse
  $t, \bar t, t$ sequences with $t$ and $\bar t$ satisfying \eqref{ttbar} also
  appear in the 19-pulse sequence found numerically in
  Ref.~\onlinecite{divincenzo00}. Based on the geometric intuition provided by
  our construction it is possible to show that both short and long $t, \bar t,
  t$ sequences can be used in this 19-pulse sequence without changing the
  resulting two-qubit gate in the total spin-1 sector (up to single qubit
  rotations).}\BibitemShut {Stop}%
\bibitem [{pro()}]{proof_note}%
  \BibitemOpen
  \href@noop {} {}\bibinfo {note} {This proof is closely related to that given
  in Ref.~\onlinecite{hormozi07} showing that for so-called $SU(2)_k$ anyons
  for $k>3$ it is necessary to braid at least five anyons to carry out similar
  two-qubit gates on qubits encoded using three anyons, and can be viewed as
  the $k\rightarrow\infty$ limit of this proof.}\BibitemShut {Stop}%
\end{thebibliography}

\end{document}